\documentclass[10pt,twocolumn,letterpaper]{article}

\usepackage[pagenumbers]{cvpr} 
\usepackage{graphicx}
\usepackage{amsmath}
\usepackage{amssymb}
\usepackage{booktabs}

\usepackage{array}
\usepackage{caption}
\usepackage{subcaption}
\usepackage{stackengine}
\usepackage{wrapfig}

\usepackage[table,xcdraw]{xcolor}

\usepackage[export]{adjustbox}
\usepackage{booktabs}
\usepackage{multirow}
\usepackage{makecell}
\usepackage{algorithm}
\usepackage{algpseudocode}
\usepackage{url}
\usepackage{tikz}
\usetikzlibrary{calc,spy}
\usepackage{cancel}
\usepackage{diagbox}
\usepackage{pifont}
\usepackage[accsupp]{axessibility}

\newcommand*\samethanks[1][\value{footnote}]{\footnotemark[#1]}

\definecolor{cvprblue}{rgb}{0.21,0.49,0.74}
\usepackage[pagebackref,breaklinks,colorlinks,citecolor=cvprblue]{hyperref}

\usepackage[capitalize]{cleveref}
\crefname{section}{Sec.}{Secs.}
\Crefname{section}{Section}{Sections}
\Crefname{table}{Table}{Tables}
\crefname{table}{Tab.}{Tabs.}

\def\paperID{10200} 
\def\confName{CVPR}
\def\confYear{2024}
\title{JDEC: JPEG Decoding via Enhanced Continuous Cosine Coefficients}
\author{
Woo Kyoung Han$^{1,2}$ \quad\qquad Sunghoon Im$^2$ \quad\qquad Jaedeok Kim$^3$\thanks{Corresponding author.} \quad\qquad Kyong Hwan Jin$^1$\samethanks\\
$^1$Korea University \quad\qquad $^2$DGIST \quad\qquad $^3$NVIDIA\\
{\tt\small \{wookyoung0727, kyong\_jin\}@korea.ac.kr, sunghoonim@dgist.ac.kr, jaedeokk@nvidia.com}
}

\newcommand{\xmark}{\ding{55}}%
\begin{document}

\maketitle
\begin{abstract}

We propose a practical approach to JPEG image decoding, utilizing a local implicit neural representation with continuous cosine formulation. The JPEG algorithm significantly quantizes discrete cosine transform (DCT) spectra to achieve a high compression rate, inevitably resulting in quality degradation while encoding an image.  We have designed a continuous cosine spectrum estimator to address the quality degradation issue that restores the distorted spectrum. By leveraging local DCT formulations, our network has the privilege to exploit dequantization and upsampling simultaneously. Our proposed model enables decoding compressed images directly across different quality factors using a single pre-trained model without relying on a conventional JPEG decoder. As a result, our proposed network achieves state-of-the-art performance in flexible color image JPEG artifact removal tasks. Our source code is available at \url{https://github.com/WooKyoungHan/JDEC}.


\end{abstract}

\section{Introduction}\label{sec:intro}

\ \ \ Within the dynamic evolution of high-efficiency image compression, it is notable that JPEG \cite{jpegstandard} maintains a pivotal position. JPEG, renowned for its compatibility and standardization, is the most famous image coder-decoder (CODEC) among conventional lossy compression methods. Therefore, a high-quality JPEG decoder applies to all existing compressed JPEG files. JPEG reduces file size through downsampling color components and quantizing the discrete cosine transform (DCT) spectra, which leads to a complicated loss of image information and distortion. Consequently, the design of a high-quality JPEG decoder presents a dual challenge: 1) the restoration of complex losses from the JPEG encoder and 2) the modeling of a network that employs a spectrum as an input and its image as an output.

 
\begin{figure}[t]
\centering
\includegraphics[trim = 340 130 270 172,clip, width = 3.2in]{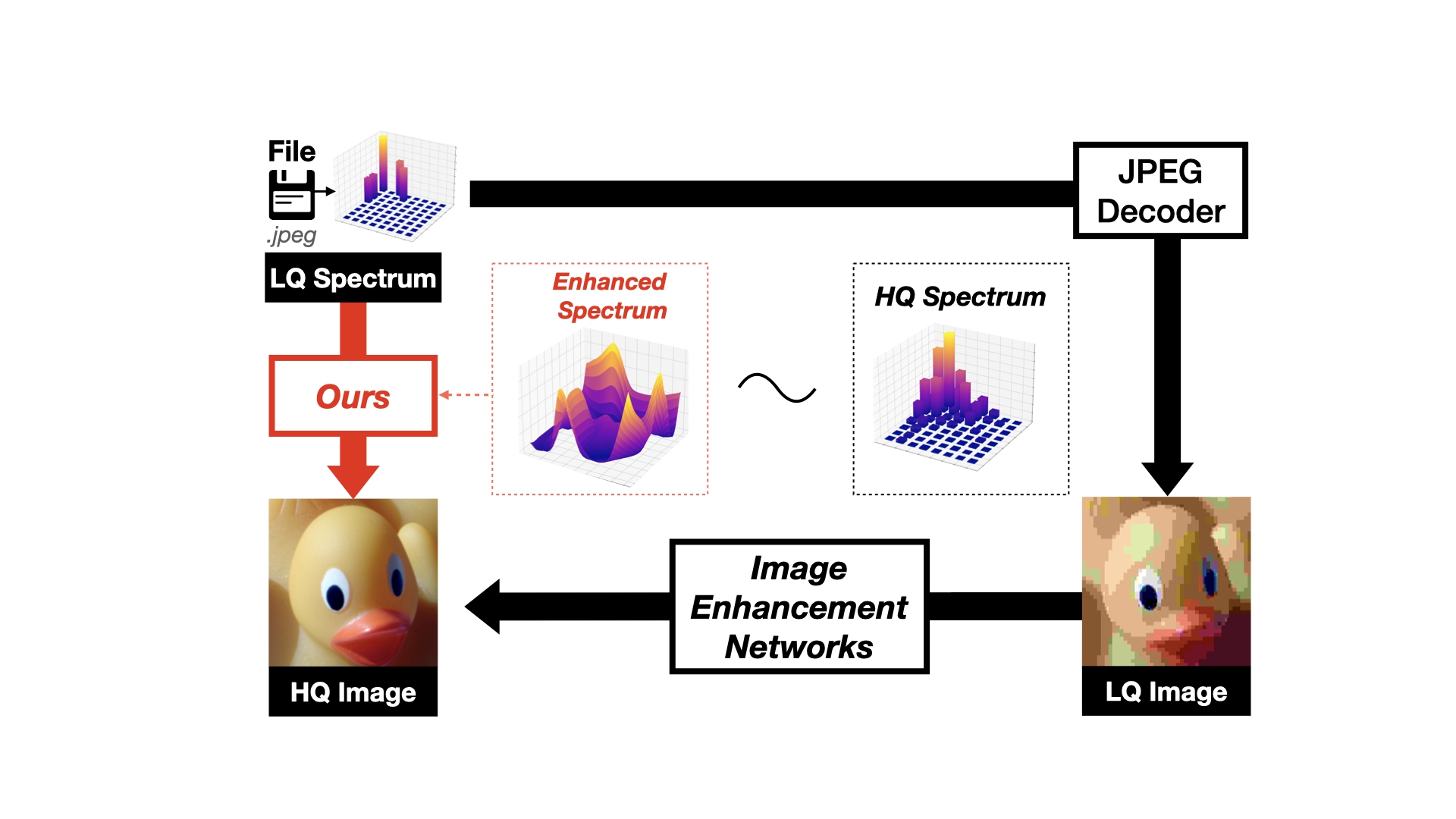}
\vspace{-8pt}
\caption{\textbf{Overall concept of proposed JPEG decoding} Instead of using a conventional JPEG decoder to refine the high-quality (HQ) image from the low-quality (LQ) image, our JDEC directly decodes the LQ spectrum by learning a continuous spectrum.}
\label{fig:concept}
\vspace{-15pt}
\end{figure}  



Many deep neural networks (DNNs) have been proposed as promising solutions for the JPEG artifact removal \cite{arcnn,mwcnn,fbcnn,qgac,dncnn,idcn,dmcnn}. Most existing methods, such as \cite{arcnn,mwcnn}, are dedicated to specific quality factors, providing multiple models to cover JPEG compression. 
In recent studies \cite{qgac,fbcnn}, the quality-dedicated problem has been addressed through the utilization of quantization maps \cite{qgac} or the estimation of quality factors \cite{fbcnn}. 
The existing artifact removal networks commonly take the decoded image as input, even though the encoded spectrum contains more information than the decoded image, according to the data processing inequality \cite{informationtheorybook}. The property is explained in \cref{sec:supp_dpi}.


Due to the characteristics of the JPEG algorithm, it is non-trivial to design a neural network that takes spectra as inputs \cite{related_taking_freq_2018_fasterneural,relatied_taking_freq_xu2020learning}. Park \etal \cite{rgbnomore} proposed a method of processing spectra to transformers.
In the context of spectral processing, our approach extends beyond proposed classification networks \cite{related_taking_freq_2018_fasterneural,relatied_taking_freq_xu2020learning,rgbnomore} by leveraging the capabilities of the embedding strategy, paving the way for more effective decoding. The spectrum conversion aligns with recent advancements in implicit neural representation (INR), where methods adopting sinusoidal functions  \cite{Benbarka_2022_WACV,mildenhall2020nerf,sitzmann2019siren,lee2021local,lee2022learning,abcd} have demonstrated significant advancement across various tasks.

\begin{figure}[t]
\footnotesize
\centering
\stackunder[2pt]{JPEG $(q=100)$}{\includegraphics[trim=0 85 165 25,clip,width = 1.05in]{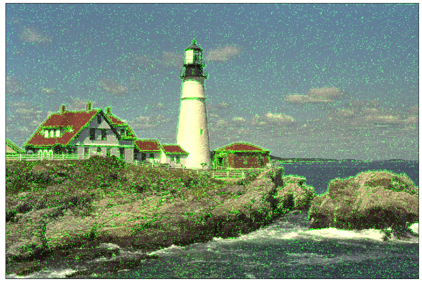}}
\hspace{-1mm}
\stackunder[2pt]{FBCNN \cite{fbcnn}}{\includegraphics[trim=0 85 165 25,clip,width = 1.05in]{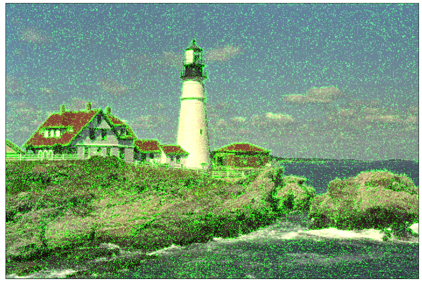}}
\hspace{-1mm}
\stackunder[2pt]{\textbf{JDEC (\textit{ours})}}{\includegraphics[trim=0 85 165 25,clip,width = 1.05in]{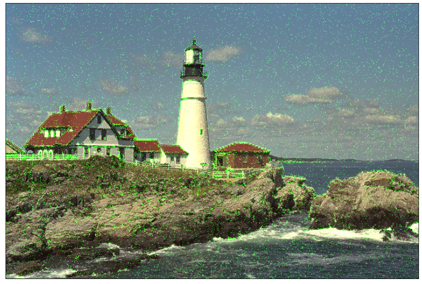}} \\
\hspace{-1mm}
{45.31 (dB) / 0.146 \qquad  43.87 (dB) / 0.175 \qquad  {\color{black}\textbf{46.80 (dB) / 0.139}}}
\vspace*{-7pt}
\caption{\textbf{Visual Demonstration at $q=100$} (PSNR (dB) $\uparrow$ / Bit-Error-Rate (BER) $\downarrow$) of decoding compressed image: JPEG (quality factor = 100), image enhancement approach \cite{fbcnn} predicted from JPEG image ($q=100$), and JDEC (\textit{ours}) predicted directly from a JPEG bit-stream. We highlight the occurrence of bit errors overlaid with green dots. }
\vspace*{-15pt}
\label{fig:visual_demo}
\vspace{-4pt}
\end{figure}

In this paper, we propose an advanced model, the JPEG Decoder with Enhanced Continuous cosine coefficients (JDEC), for retrieving high-quality images from compressed spectra. As an artifact removal network, our JDEC does not require a conventional JPEG decoder compared to existing methods shown in \cref{fig:concept}. JDEC captures the dominant frequency and its amplitude, thereby representing the high-quality spectrum through continuous cosine formulation (CCF). The CCF module estimates a continuous form of a given discrete cosine spectrum. The proposed model represents a considerable improvement in decoding JPEG bitstream. As shown in \cref{fig:visual_demo}, our JDEC decodes high-quality images with fewer bit errors than the original JPEG decoder.



In summary, our main contributions are as follows:
\begin{itemize}
    \item We propose a local implicit neural representation that decodes JPEG files across various quality factors (QF) with continuous cosine spectra.
    
    \item We show that the suggested continuous cosine formulation module lets the network predict spectra highly correlated with the ground truth's spectrum.

    \item We demonstrate that our proposed method operates as a practical decoder, delivering superior image quality, including the generally used quality factor.

    

\end{itemize}

\section{Related Work}\label{sec:related}

\begin{figure}[t]
\footnotesize
\centering
\includegraphics[width = 3.3in]{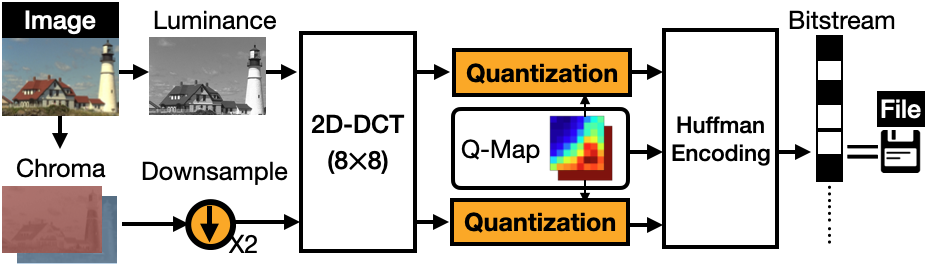}
\caption{\textbf{Overall process of the JPEG encoder.} Luminance and chroma components are separated from an RGB image. Both components are converted to DCT spectra and quantized with a pre-defined quantization matrix (Q-map). All losses occur in the orange area.}
\vspace*{-15pt}
\label{fig:jpeg_process}
\vspace{-4pt}
\end{figure}

\textbf{JPEG Background} According to Shannon's source coding theorem \cite{shannoncapacity}, a loss of image information is unavoidable to achieve high-efficiency compression.
The JPEG initiates the encoding process by decomposing an input RGB image to luminance and chroma components \cite{jpegstandard}. 
The chroma components are downsampled using the nearest neighbor method by a factor of $\times 2$. 
The JPEG subtracts the midpoint of the pixel value (=128) to images and divides it into $8\times8$ crops. Then, each crop is transformed into 2D-DCT \cite{ahmed1974discrete} spectra.
Following this, the encoder quantizes the spectrum of each block using a predefined quantization matrix depending on a quality factor $q$, and then the quantized spectrum is coded using Huffman coding. 
We illustrate the process of the JPEG encoder in \cref{fig:jpeg_process}.

Due to the nature of DCT, the energy of spectra is concentrated in low-frequency components. 
Since the quantization matrix treats high-frequency components more severely than low-frequency components, most distortions occur in the high-frequency components.
In the JPEG decoder, the quantization matrix is directly applied to the quantized spectra, transforming them into images. 
Consequently, all incurred losses, especially those in high-frequency components, are directly conveyed to the resulting image.

\noindent{\bf JPEG Artifact Removal} 
To address the aforementioned problem, learning-based methods have enhanced the quality of a decoded image. 
Dong \etal \cite{arcnn} introduced a neural network that utilizes a super-resolution network \cite{srcnn} for JPEG artifact removal. 
Most of the proposed neural networks are dedicated to a specific quality factor \cite{cas-cnn,jpegrelateddedicated,arcnn,jpegrelatedwaveletcnn}. 
To tackle the quality-dedicated issue, Jiang \etal \cite{fbcnn} proposed a method to estimate a quality factor, solving flexible JPEG artifact removal and handling a double JPEG artifact. 
However, the existing artifact removal methods take images as input, incorporating the conventional JPEG decoder before using their network. 
Recently, Bahat \etal \cite{bahat2021whatsin} proposed a novel method for JPEG decoding, which takes spectra as input. However, the proposed method does not consider color components with trainable decoding and does not recover high-quality factors.
\begin{figure*}[t]
\centering
\includegraphics[trim= 0 27 14 86,clip,width=7in]{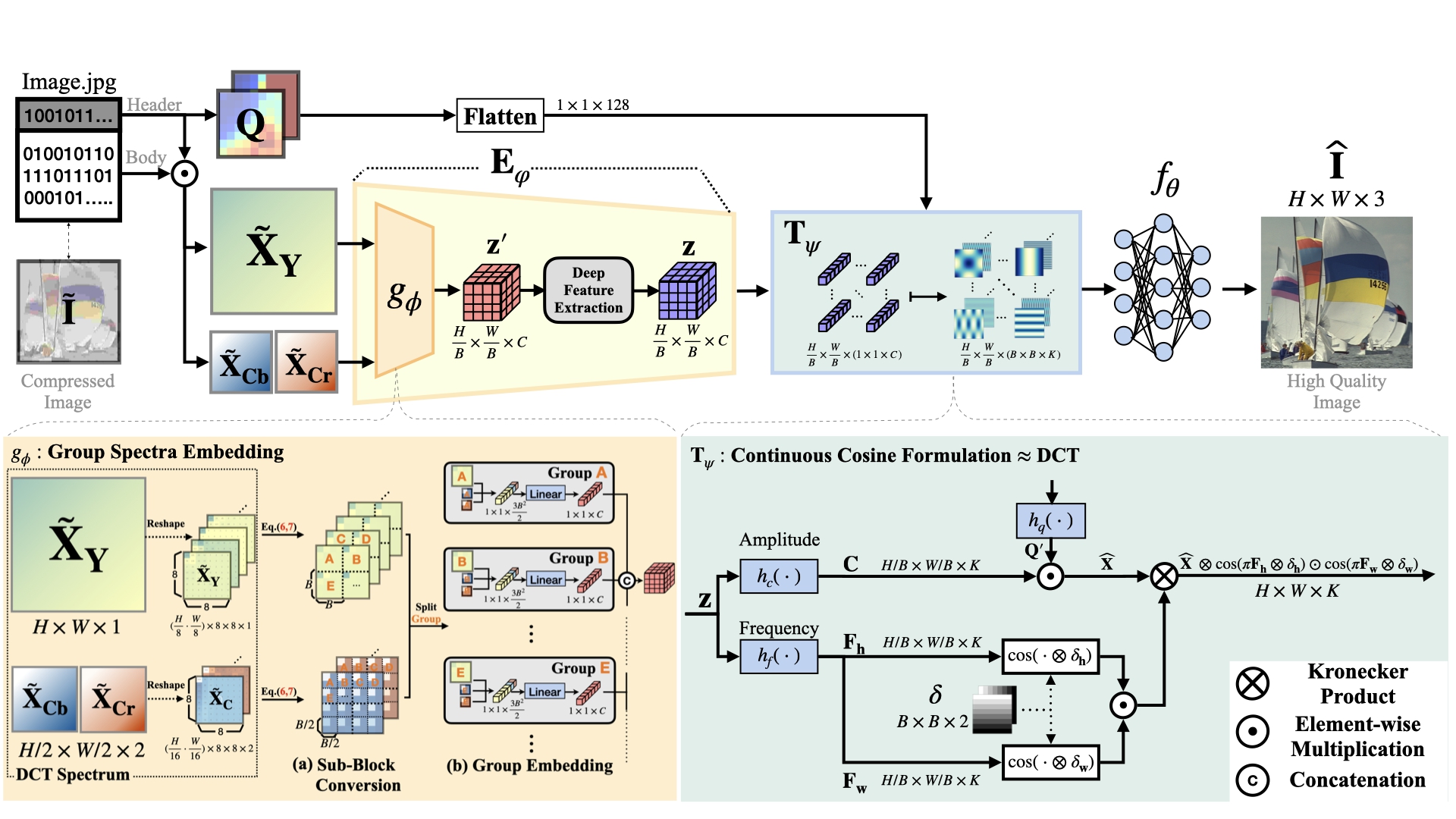}
\caption{\textbf{Decoding a JPEG bitstream with the proposed JDEC.} JDEC consists of an encoder ($E_\varphi$) with group spectra embedding ($g_\phi$), a decoder ($f_\theta$), and continuous cosine formulation ($T_\psi$). Inputs of JDEC are as follows: compressed spectra ($\mathbf{\tilde{X}_Y},{\mathbf{\tilde{X}_C}}$), quantization map $\mathbf{Q}$. Note that our JDEC does not take ${\mathbf{\tilde{I}}}$ as an input. JDEC formulates latent features into a trainable continuous cosine coefficient as a function of block grid $\delta$ and forward to INR ($f_\theta$). Therefore, each $B\times B$ block shares the estimated continuous cosine spectrum. }
\label{Figmain}
\vspace{-12pt}
\end{figure*}

\noindent{\bf Learning in the Frequency domain} 
 In an image classification task, skipping a conventional JPEG decoding \cite{related_taking_freq_2018_fasterneural,relatied_taking_freq_xu2020learning} has been proposed, especially optimizing CNNs. Embedding techniques \cite{related_taking_freq_2018_fasterneural} tackle the size mismatch issue between luma and chroma components, such as upsampling chroma components before forwarding to a network and upsampling chroma features after forwarding to a shallow network. The proposed methods boost computation time without dropping the original performance. Recently, the approach adopting vision transformers \cite{vit2020ICLR} instead of CNNs has promising performance \cite{rgbnomore}. We adopt the proposed embedding method from \cite{rgbnomore} and modified \cite{Liu_2022_CVPR_swin2} SwinV2 transformer suitable for image decoding.

\section{Method}\label{problemformulation}



\noindent\textbf{Problem Formulation}  Let $\mathbf{I_{GT}} \in \mathbb{R}^{H\times W \times 3}$ be a ground-truth RGB image. 
The JPEG encoder separates $\mathbf{I_{GT}}$ to luminance component ($\mathbf{I_{Y}}\in \mathbb{R}^{H\times W \times 1}$) and chroma components ($\mathbf{I_{C}}\in \mathbb{R}^{H\times W \times 2}$) and downsamples chroma components by a factor of 2, i.e. ($\mathbf{{I}^{\downarrow}_{C}}\in \mathbb{R}^{\frac{H}{2}\times \frac{W}{2} \times 2}$). The superscript $\downarrow$ indicates a $\times 2$ downsampling. Then, each component is divided into $8\times8$ blocks ($\mathbf{I} \in \mathbb{R}^{8\times8} \subset \mathbf{I_{Y}},\mathbf{{I}^{\downarrow}_{C}}$). 2D-DCT \cite{ahmed1974discrete} into spectra $\mathbf{X}\in\mathbb{R}^{8\times8}  $ is defined as below:
\begin{align}\label{eq:dct}
   DCT(\mathbf{I}) := \mathbf{X} &= \mathbf{D}\mathbf{I}\mathbf{D}^\top.
\end{align}
The orthonormal basis matrix $\mathbf{D}(=\mathbf{D}_{8})$ is defined as:
\begin{align}
      \tiny\mathbf{D}_N&:= \left[ \alpha \right] \odot \cos([\pi {[{F_{k|N}}]^{N-1}_{k=0}}\otimes [k]^{N-1^\top}_{k=0}])\\
      =&\sqrt{\frac{2}{N}}\left[\begin{smallmatrix}
         \frac{1}{\sqrt{2}}                & \frac{1}{\sqrt{2}} &\cdots  & \frac{1}{\sqrt{2}}\\
         \cos \left(\frac{1\pi}{2N}\right) &\cos \left(\frac{3\pi}{2N}\right) &\cdots &\cos \left(\frac{(2N-1)\pi}{2N}\right)\\
            \vdots &\vdots &\ddots &\vdots\\
         \cos \left(\frac{(N-1)\pi}{2N}\right) &\cos \left(\frac{3(N-1)\pi}{2N}\right) &\cdots &\cos \left(\frac{(2N-1)(N-1)\pi}{2N}\right)\\ \nonumber
     \end{smallmatrix}\right].\label{eq:orthobasis}
\end{align}
where ${F_{k|N}} := (2k+1) \slash 2N$ is a fixed frequency of a coordinate $k$ with a given size $N$ and $[\alpha]$ is the scaling matrix for orthonormality.
The operations $\otimes$ and $\odot$ are a Kronecker product and element-wise multiplication, respectively.

Quantization is conducted with a predefined quantization matrix $\mathbf{Q} = [\mathbf{{Q}_{Y}};\mathbf{Q_C}]\in {\mathbb{N}^{8\times8\times2}}$ s.t.
\vspace*{-5pt}
\begin{equation}\label{eq:quantize_spectra}
    \tilde{\mathbf{C}} := \left\lfloor  \mathbf{X}\odot \frac{1}{\mathbf{Q}}  \right\rceil,~
    \Tilde{\mathbf{X}} =  \tilde{\mathbf{C}}  \odot \mathbf{Q},
\end{equation}
where $\lfloor \cdot \rceil$ is a round operation that maps to the nearest integer. $1/\mathbf{Q}$ denotes an element-wise division. 
The JPEG encoder compresses the header $\mathbf{Q}$ and the body of a code $\tilde{\mathbf{C}}$ separately. 
In the decoder part, the JPEG restores the image from the compressed header and body by $\tilde{\mathbf{I}} = DCT^{-1}(\Tilde{\mathbf{C}} \odot \mathbf{Q})$.

 To summarize this, the corrupted JPEG image $\tilde{\mathbf{I}}$ is obtained by
\begin{align}\label{eq:overall}
 \begin{bmatrix}
    \mathbf{\Tilde{I}_{Y}} \\
   \mathbf{\Tilde{I}_{C}}
 \end{bmatrix}= \begin{bmatrix}
   DCT^{-1}(\lfloor DCT(\mathbf{I_{Y}})\odot\frac{1}{{\mathbf{Q}_Y}}\rceil\odot\mathbf{Q_Y})\\
   DCT^{-1}(\lfloor DCT(\mathbf{I^{\downarrow}_{C}})\odot\frac{1}{{\mathbf{Q}_C}}\rceil\odot\mathbf{Q_C})^{\uparrow}
 \end{bmatrix},
\end{align}
where 
the superscript $\uparrow$ indicates $\times2$ upsampling in the spatial domain. 
We here observe from \cref{eq:overall} that most of the loss of image information is induced by the quantization step.
The shape of the chroma component $\mathbf{\tilde{{X}}_{C}}\in \mathbb{R}^{\frac{H}{2}\times\frac{W}{2}\times2}$ is different from the luminance component $\mathbf{\tilde{X}_Y} \in \mathbb{R}^{{H}\times{W}\times 1}$. 
We will consider observations in the design of our proposed network.




We propose a JPEG decoder network, {JDEC} $\mathbf{J_\Theta}$, defined by
\begin{align}
    \mathbf{J_{\Theta}} \colon 
        (\mathbf{\tilde{X}_{Y},\tilde{X}_{C};Q}) \mapsto \mathbf{\widehat{I}}.
\label{eq:jdecoveall}
\end{align}
The network directly accepts quantized spectrum $\mathbf{\tilde{X}_{Y}}$ and $\mathbf{\tilde{X}_{C}}$ with a quantization matrix as the network inputs, enabling to decode JPEG directly from encoded JPEG data.
Our proposed JDEC comprises mainly three parts: encoder $\mathbf{E}_{\varphi}$ with group embedding, continuous cosine formulation $\mathbf{T}_{\psi}$, and decoder $f_\theta$ with an implicit neural representation.



\noindent\textbf{Encoder ($\mathbf{E_\varphi}$)} 
The encoder is a function $\mathbf{E}_\varphi \colon (\mathbf{\tilde{X}_Y} , \mathbf{\tilde{X}_C}) \mapsto \mathbf{z} \in \mathbb{R}^{ \frac{H}{B}\times \frac{W}{B}\times C}$. 
We model the encoder $\mathbf{E}_\varphi$ by using SwinV2 \cite{Liu_2022_CVPR_swin2}.
To allow the different shape of spectrum $\mathbf{\tilde{X}_Y}$ and $\mathbf{\tilde{X}_C}$,
we apply the group spectra embedding layer $g_\phi$ proposed by \cite{rgbnomore}. The embedding layer ($g_\phi$) transforms luminance and chroma spectra through two steps. We convert $8\times8$ spectra into $B\times B$ for luma and $B/2\times B/2$ for chroma via sub-block conversion \cite{subblockconversion} in the (a) part of $g_\phi$ in \cref{Figmain}. We implement the block size $B = 4$\footnote{$B$ should be divisor or multiple of 16}. 
\vspace{-1pt}
\begin{align}
    \mathbf{X_{Y}'} &= \mathbf{D}^{*}_{B}\mathbf{(D^\top X_{Y}D)D}_{B}^{*\top}, \\
    \mathbf{X_{C}'} &= \mathbf{D}^{*}_{B/2}\mathbf{(D^\top X_{C}D)D}_{B/2}^{*\top},
\text{\vspace{-10pt}}
\end{align}
where $D^{*}_N$ indicates a block diagonal matrix with size $8\times8$.
In part (b) of \cref{Figmain}, spectra are reshaped and concatenated to $\mathbb{R}^{\frac{H}{B} \times \frac{W}{B} \times \frac{3B^2}{2}}$ which is the sum of converted size. Then initialized latent vector $\mathbf{z'} \in \mathbb{R}^{\frac{H}{B} \times \frac{W}{B} \times C}$ are conducted in (b) part of $g_\phi$. Following the prior work, \cite{liang2021swinir}, we adopt the deep feature extractor, replacing the Swin attention module with the SwinV2 attention module.





\begin{figure}[t]
\footnotesize
\centering
\includegraphics[trim= 180 0 0 540,clip,width = 3.3in]{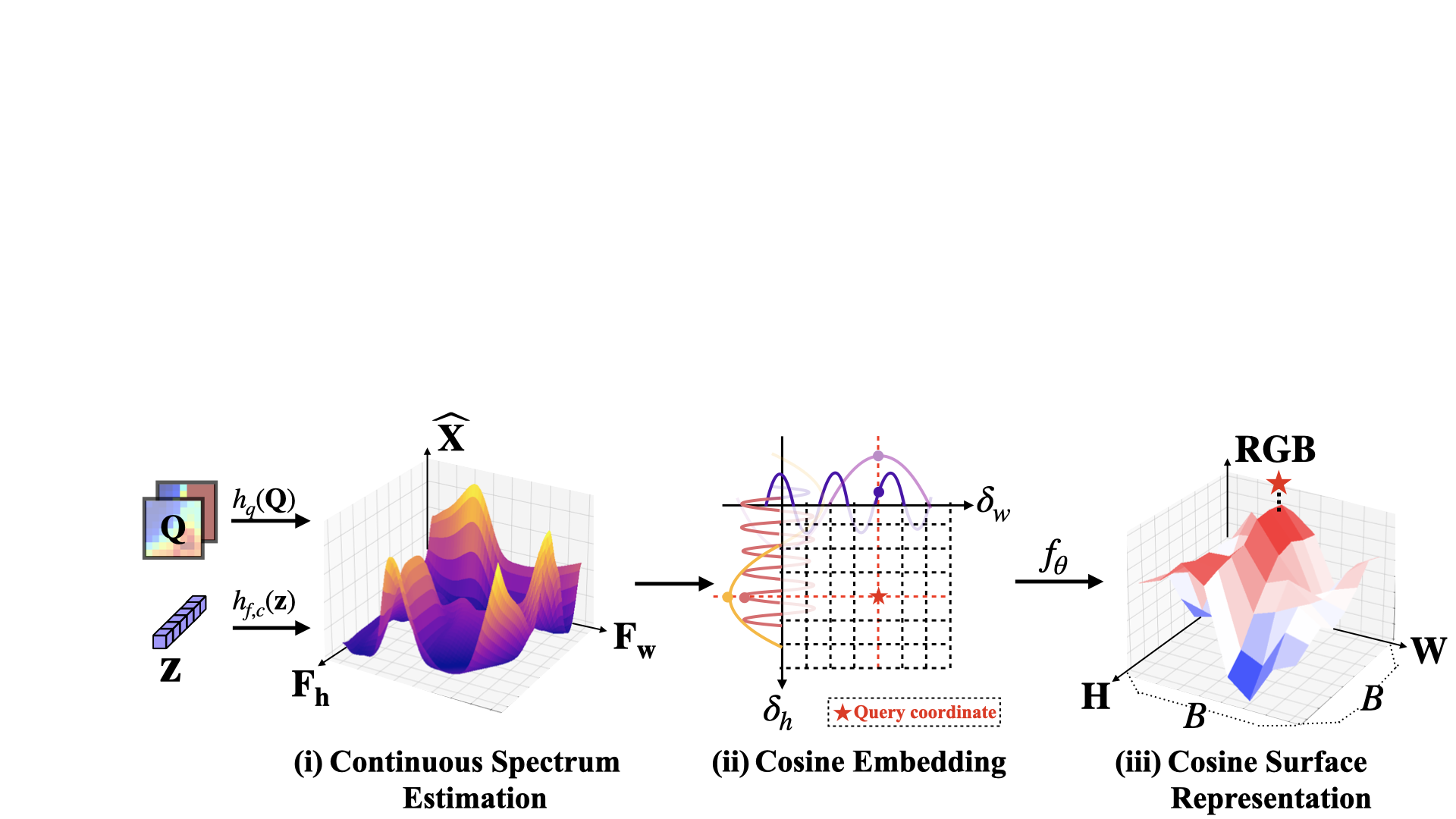}
\vspace{-17pt}
\caption{\textbf{Graphical summary of $f_\theta( T_\psi (\delta, \mathbf{z} ;\mathbf{Q}))$.} Each $1\times1$-sized feature $\mathbf{z}$ maps into a $B\times B$ pixel area. $T_\psi$ embeds the local coordinates of $B\times B$ area and forwards to $f_\theta$.}
\label{fig:graphical}
\vspace{-15pt}
\end{figure}
 
\noindent\textbf{Continuous Cosine Formulation ($\mathbf{T}_\psi$)}
Each JPEG block shares a distorted DCT spectrum.
Modifying the entire spectrum is required to restore the distortion of a block.
To address the spectrum distortion issue derived from the JPEG encoder, we introduce \textit{Continuous Cosine Formulation} (CCF) module, which enhances the cosine spectrum. The CCF constructs a continuous spectrum corresponding to $B\times B$ embedded block by estimating dominant frequencies and amplitudes of a cosine transform. Illustrated in Figure \ref{fig:graphical}, each block has identical amplitudes and frequencies within the embedded block coordinate $\delta :=\left[(i,j)\right]^{B-1}_{i,j=0} $. 


Our CCF takes a latent vector $\mathbf{z}$ from encoder $\mathbf{E_\varphi}$ and a quantization matrix $\mathbf{Q}$. The CCF $\mathbf{T}_\psi$ consists of three elements: frequency estimator $h_f \colon \mathbb{R}^C\mapsto \mathbb{R}^{2K}$, coefficient estimator $h_c \colon \mathbb{R}^C\mapsto \mathbb{R}^K$, and quantization matrix encoder $h_q \colon \mathbb{R}^{128} \mapsto \mathbb{R}^K$. Each frequency and coefficient estimator comprises sequential convolution and non-linear activation layers.
As a method for quantization recovery, we implement an amplitude recovery method as described below, drawing inspiration from the existing dequantization network \cite{abcd}, 
\begin{align}
     \mathbf{\widehat{X} =C \odot Q'} \sim \cref{eq:quantize_spectra},
\end{align}  
where $\mathbf{Q}' = h_q(\mathbf{Q})$ and $\mathbf{C} = h_c(\mathbf{z})$.

We hypothesize that estimating the frequency components effectively mitigates aliasing (i.e., quantization and downsampling) derived from JPEG.
It has been demonstrated that trainable frequencies and phasors effectively mitigate upsampling and dequantization \cite{abcd,lee2022learning}. 

We thus formulate the CCF module approximates $B\times B$ spectral features from the fiber of $\mathbf{z}$:
\begin{align}
    \mathbf{T}_\psi(\mathbf{z},\delta_{\mathbf{h,w}};\mathbf{Q}) =& \mathbf{\widehat{X}\otimes(\cos(\pi\mathbf{F}_{h}\otimes \delta_h)\odot \cos(\pi \mathbf{F}_{w}\otimes\delta_w))},
\end{align} 
where $[\mathbf{F_h};\mathbf{F_w}]= h_f(\mathbf{z})$. $\mathbf{\delta_{h,w}}$ denotes vertical and horizontal coordinates of $\delta$. Note that $\mathbf{\widehat{X}}, \mathbf{F_{h}},\mathbf{F_{w}} \in \mathbb{R}^{K}$ are amplitude and frequencies for the spatial coordinate $\delta$, respectively. i.e. the CCF maps embedded features and block coordinates by $\mathbf{T_\psi} : (\mathbb{R}^{1\times1\times{C}},\mathbb{R}^{B\times B\times{2}})\mapsto \mathbb{R}^{B\times B\times{K}}$.


\noindent\textbf{Decoder ($f_\theta$)} Our decoder $f_\theta\colon \mathbb{R}^{K}\mapsto \mathbb{R}^3$ is a local implicit neural representation function of $\{\mathbf{z},\mathbf{Q}\},\delta$. i.e. :
\begin{align}\label{eq:inr_final}
\tiny
    \mathbf{\widehat{I}} &= f_\theta(\mathbf{\widehat{X}\otimes(\cos(\pi\mathbf{F}_{h}\otimes\delta_h)\odot \cos(\pi \mathbf{F}_{w}\otimes\delta_w))}).
\end{align}
Therefore, in the $B\times B$ block of \cref{eq:inr_final}, the estimated basis of $\mathbf{\hat{X}}$ and its reconstruction follows:
\begin{align}
 \mathbf{I} &= \mathbf{D^\top XD} \simeq f_\theta'(\mathbf{\Lambda_h}\mathbf{\widehat{X}'}\mathbf{\Lambda_w}),
 \vspace{-8pt}
\end{align}
where $\mathbf{\Lambda_{h,w}} = \cos([\pi \mathbf{F_{h,w}}\otimes{\mathbf{\delta_{h,w}}}])$ and $f'_\theta$ satisfy $f_\theta = f'_\theta \circ W$ for a trainable fully-connected layer $W$. With a linear layer $W$, the \cref{eq:inr_final} complete the quadratic form $\mathbf{\Lambda}_1\mathbf{\widehat{X}'}\mathbf{\Lambda}_2 = W(\mathbf{T_\psi(z,Q;\delta)})$ by including summation of features. 
We optimize a set of trainable parameters $\Theta := \{\varphi ;\psi;\theta \}$ with the equation below:
\begin{equation}
      \widehat\Theta = \arg\min_{\Theta} || \mathbf{I}_{GT} - \widehat{\mathbf{I}}(\mathbf{\tilde{X}_Y,\tilde{X}_C,Q} ; \Theta) ||_{1}.
  \end{equation}
  
We will demonstrate the estimated frequencies ($\mathbf{F_h},\mathbf{F_w}$) and amplitudes $\mathbf{\widehat{X}}$ of networks follows $\mathbf{X}$ in the following section.

\begin{figure*}[t]
    \centering
    \includegraphics[trim=0 0 0 0,clip,width = 3.3in]{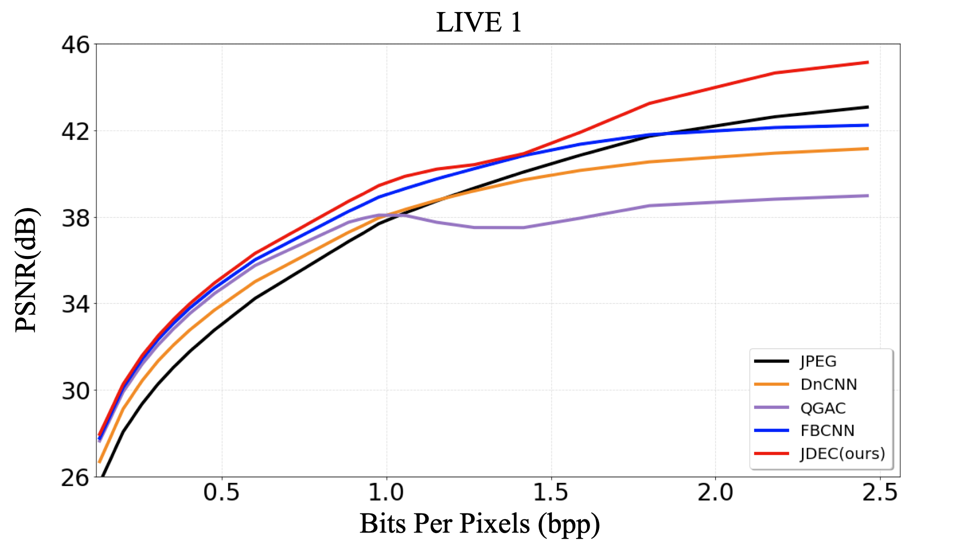} \includegraphics[trim=0 0 0 0,clip,width = 3.3in]{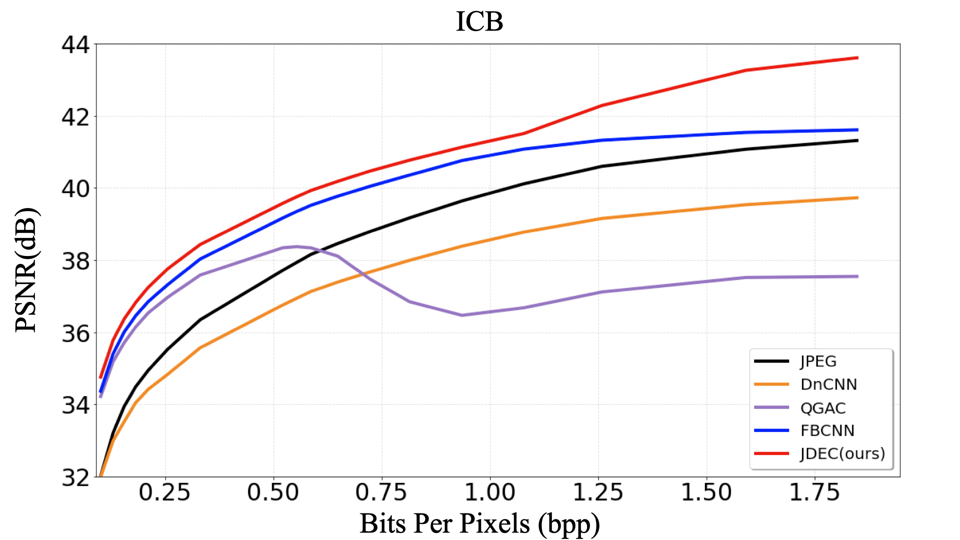}
    \includegraphics[trim=0 0 0 0,clip,width = 3.3in]{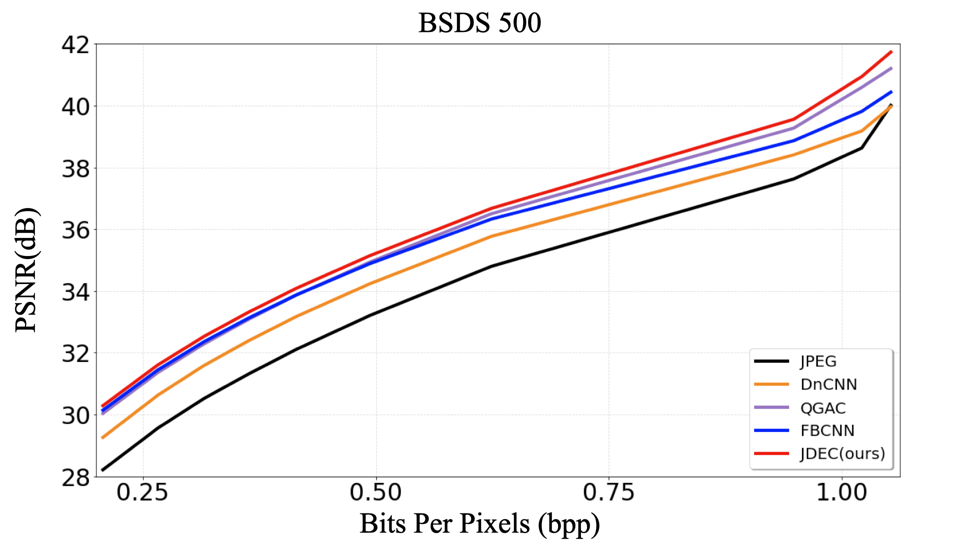}
    \includegraphics[trim=0 0 0 0,clip,width = 3.3in]{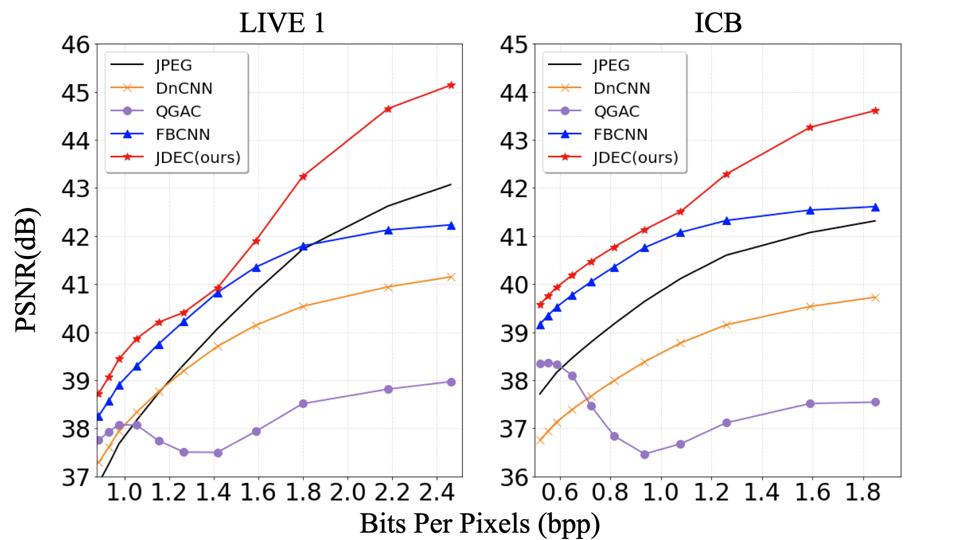}
    \vspace{-10pt}
    \caption{RD curve results on LIVE-1 \cite{Live1} (top left), ICB \cite{ICBdataset} (top right), BSDS500 \cite{B500dataset} (bottom left). We highlight the high-quality factor parts $q\in[90,100]$ in the bottom right part. We show PSNR as a measure of \textit{distortion} (higher is better). We observe that our JDEC decodes high-quality images better than other methods.}

    \label{fig:RDcurve-qual}
\end{figure*}

\begin{table*}[t]
\centering
\setlength{\tabcolsep}{1.2pt}
\scriptsize{
\begin{tabular}{c|>{\centering\arraybackslash}p{1.23cm}>{\centering\arraybackslash}p{1.23cm}>{\centering\arraybackslash}p{1.23cm}>{\centering\arraybackslash}p{1.23cm}
|>{\centering\arraybackslash}p{1.23cm}>{\centering\arraybackslash}p{1.23cm}>{\centering\arraybackslash}p{1.23cm}>{\centering\arraybackslash}p{1.23cm}
|>{\centering\arraybackslash}p{1.23cm}>{\centering\arraybackslash}p{1.23cm}>{\centering\arraybackslash}p{1.23cm}>{\centering\arraybackslash}p{1.23cm}}
Test & \multicolumn{4}{c|}{LIVE-1 \cite{Live1}} &\multicolumn{4}{c|}{BSDS500 \cite{B500dataset}} &\multicolumn{4}{c}{ICB \cite{ICBdataset}}\\
\cline{1-13}
Method & $q =10$ & $q=20$& $q=30$ & $q=40$ &  $q=10$ & $q=20$& $q=30$ & $q=40$ &  $q=10$ & $q=20$& $q=30$ & $q=40$ \\
\cline{1-2}
\hline\hline

\multirow{2}{*}{JPEG}  
& 25.69$|$24.20 & 28.06$|$26.49 & 29.37$|$27.84 & 30.28$|$28.84 
& 25.84$|$24.13 & 28.21$|$26.37 & 29.57$|$27.72 & 30.52$|$28.69
& 29.44$|$28.53 & 32.01$|$31.11 & 33.20$|$32.35 & 33.95$|$33.14 \\
&    0.759      &    0.841      &    0.875      &    0.894 
&    0.759      &    0.844      &    0.880      &    0.900 
&    0.753      &    0.807      &    0.833      &    0.844    \\
\multirow{2}{*}{DMCNN \cite{dmcnn}}  
& 27.18$|$27.03 & 29.45$|$29.08 & - & - 
& 27.16$|$26.95 & 29.35$|$28.84 & - & - 
& 30.85$|$\quad  -\quad\quad\quad & 32.77$|$\quad  -\quad\quad\quad & - & -              \\
&    0.810      &    0.874      & - & -  
&    0.799      &    0.866      & - & -  
&    0.796      &    0.830      & - & -            \\
\multirow{2}{*}{IDCN \cite{idcn}}  
& 27.62$|$27.32 & 30.01$|$29.49 & - & - 
& 27.61$|$27.22 & 28.01$|$25.57 & - & -
& 31.71$|$\quad  -\quad\quad\quad& 33.99$|$\quad  -\quad\quad\quad  & - & - \\
&    0.816     &    0.881    &    -     &    - 
&    0.805     &    0.873    &    -     &   - 
&    0.809     &   0.838    &    -     &    -\\
\multirow{2}{*}{Swin2SR* \cite{conde2022swin2sr} }
& \textcolor{red}{27.98}$|$\quad  -\quad\quad\quad   &   -     &    - & \textcolor{red}{32.53}$|$\quad  -\quad\quad\quad 
&   -    &   -     &    -      &   -
& \textcolor{blue}{32.46}$|$\quad  -\quad\quad\quad  &- & - & \textcolor{blue}{36.25}$|$\quad  -\quad\quad\quad  \\
&   -    &   -     &    -      &   -  
&   -    &   -     &    -      &   -  
&   -    &   -     &    -      &   -  \\
\hline
\multirow{2}{*}{DnCNN \cite{dncnn}}  
& 26.68$|$26.47 & 29.12$|$28.77 & 30.43$|$30.04 & 31.34$|$30.94 
& 26.82$|$26.53 & 29.26$|$28.74 & 30.63$|$30.02 & 31.59$|$30.92
& 29.78$|$29.71 & 31.99$|$31.90 & 32.98$|$32.89 & 33.52$|$33.42 \\
&    0.794      &    0.866      &    0.895      & 0.911
&    0.793      &    0.867      &    0.898      & 0.915
&    0.726      &    0.765      &    0.786      & 0.978             \\
\multirow{2}{*}{QGAC \cite{qgac}}  
& 27.65$|$27.43 & 29.88$|$29.56 & 31.17$|$30.77 & 32.08$|$31.64 
& 27.75$|$27.48 & 30.04$|$29.55 & 31.36$|$30.73 & 32.29$|$31.53
& 32.12$|$32.09 & 34.22$|$34.18 & 35.18$|$35.13 & 35.71$|$35.65 \\
&    \textcolor{blue}{0.819}      &    \textcolor{blue}{0.882}      &    0.908      &    0.922 
&    \textcolor{blue}{0.819}      &    \textcolor{blue}{0.884}      &    0.911      &    0.926 
&    \textcolor{blue}{0.814}      &    \textcolor{blue}{0.844}      &    \textcolor{blue}{0.859}      &    0.865\\
\multirow{2}{*}{FBCNN \cite{fbcnn}}  
& \textcolor{black}{27.77}$|$\textcolor{blue}{27.51} & \textcolor{blue}{30.11}$|$\textcolor{blue}{29.70} & \textcolor{blue}{31.43}$|$\textcolor{blue}{30.92} & \textcolor{black}{32.34}$|$\textcolor{blue}{31.80} 
& \textcolor{blue}{27.85}$|$\textcolor{blue}{27.53} & \textcolor{blue}{30.14}$|$\textcolor{blue}{29.58} & \textcolor{blue}{31.45}$|$\textcolor{blue}{30.74} & \textcolor{blue}{32.36}$|$\textcolor{blue}{31.54}
& \textcolor{black}{32.18}$|$\textcolor{blue}{32.15} & \textcolor{blue}{34.38}$|$\textcolor{blue}{34.34} & \textcolor{blue}{35.41}$|$\textcolor{blue}{35.35} & \textcolor{black}{36.02}$|$\textcolor{blue}{35.95} \\
&    \textcolor{black}{0.816}    &    \textcolor{black}{0.881}      &    \textcolor{blue}{0.908}      &    \textcolor{blue}{0.923}  
&    \textcolor{black}{0.814}      &    \textcolor{black}{0.881}      &    \textcolor{black}{0.909}      &    \textcolor{black}{0.924} 
&    \textcolor{black}{0.813}      &    \textcolor{blue}{0.844}      &    \textcolor{blue}{0.859}      &    \textcolor{blue}{0.869}\\
\multirow{2}{*}{\textbf{JDEC \textit{(ours)}}}  
& \textcolor{blue}{27.95}$|$\textcolor{red}{27.71} & \textcolor{red}{30.26}$|$\textcolor{red}{29.87} & \textcolor{red}{31.59}$|$\textcolor{red}{31.12} & \textcolor{blue}{32.50}$|$\textcolor{red}{31.98}
& \textcolor{red}{28.00}$|$\textcolor{red}{27.67} & \textcolor{red}{30.31}$|$\textcolor{red}{29.71} & \textcolor{red}{31.65}$|$\textcolor{red}{30.88} & \textcolor{red}{32.53}$|$\textcolor{red}{31.68} 
& \textcolor{red}{32.55}$|$\textcolor{red}{32.51} & \textcolor{red}{34.73}$|$\textcolor{red}{34.68} & \textcolor{red}{35.75}$|$\textcolor{red}{35.68} & \textcolor{red}{36.37}$|$\textcolor{red}{36.28} \\
&   \textcolor{red}{ 0.821}      &    \textcolor{red}{0.885 }     &    \textcolor{red}{0.911}      &    \textcolor{red}{0.925} 
&    \textcolor{red}{0.819}      &    \textcolor{red}{0.885}      &    \textcolor{red}{0.912}      &    \textcolor{red}{0.927} 
&   \textcolor{red}{0.818}     &    \textcolor{red}{0.847}      &    \textcolor{red}{0.862}      &    \textcolor{red}{0.871}
\end{tabular}}
\vspace*{-6pt}
\caption{Quantitative comparisons (PSNR (dB) $|$ PSNR-B (dB) (top), SSIM (bottom)) with \textit{the color JPEG artifact removal} networks. \textcolor{red}{Red} and \textcolor{blue}{blue} colors indicate the best and the second-best performance, respectively. (-) indicates not reported. (*) indicates using additional datasets. Note that only JPEG \cite{jpegstandard} and our JDEC get spectra as input.}
    \vspace{-10pt}
\label{tab:Quan_testvalid}
\end{table*}

\section{Experiments}
\subsection{Network Details}
\noindent \textbf{Encoder ($\mathbf{E}_\varphi$) and Decoder ($f_\theta$)} The linear layer in group spectra embedding module $g_\phi$ has an embedding size $C$ of 256. We modified the deep feature extract part of SwinIR \cite{liang2021swinir}. The window attention module is replaced with SwinV2 \cite{Liu_2022_CVPR_swin2}, with a window size of 7. 
\cite{liang2021swinir} and \cite{rgbnomore} reported that a window size of 8 significantly drops the performance of the network.  Each residual Swin transformer block includes 6 Swin transformer layers. The decoder $f_\theta$ is an MLP composed of 5 linear layers with 512 hidden channels $K$ and ReLU activations.

\noindent \textbf{CCF} The CCF includes a frequency estimator $h_f$, an amplitude estimator $h_c$, and a quantization matrix encoder $h_q$. 
\cite{lee2021local,lee2022learning,abcd} show that learning frequency, phase, and amplitude components enhance the performance of the INR. The quantization matrix encoder $h_q$ is a single fully connected layer, having 512($=K$) channels. The amplitude and frequency estimator ($h_c, h_f$) is designed with two $3\times3$ convolutional layers with a ReLU activation. The frequency estimator has $2K (= 1024)$ output channels for $h$ and $w$ axis, while the amplitude estimator has $K (=512)$ channels.

\tikzstyle{largewindow_w} = [white, line width=0.30mm]
\tikzstyle{smallwindow_w} = [white, line width=0.10mm]
\tikzstyle{largewindow_b} = [blue, line width=0.30mm]
\tikzstyle{smallwindow_b} = [blue, line width=0.10mm]
\tikzstyle{closeup_b} = [
  opacity=1.0,          
  height=1cm,         
  width=1cm,          
  connect spies, blue  
]
\tikzstyle{closeup_w} = [
  opacity=1.0,          
  height=1cm,         
  width=1cm,          
  connect spies, white  
]
\tikzstyle{closeup_w_2} = [
  opacity=1.0,          
  height=1.7cm,         
  width=1.7cm,          
  connect spies, white  
]
\tikzstyle{closeup_w_3} = [
  opacity=1.0,          
  height=1.35cm,         
  width=1.35cm,          
  connect spies, white  
]
\tikzstyle{closeup_w_4} = [
  opacity=1.0,          
  height=0.8cm,         
  width=0.8cm,          
  connect spies, white  
]
\tikzstyle{closeup_w_5} = [
  opacity=1.0,          
  height=0.6in,         
  width=0.6in,          
  connect spies, white  
]

\begin{figure*}[t]
\footnotesize
\centering
\hspace{-7pt}
\raisebox{0.2in}{\rotatebox{90}{BSDS500 \cite{B500dataset}}}
\begin{tikzpicture}[x=6cm, y=6cm, spy using outlines={every spy on node/.append style={smallwindow_w}}]
\node[anchor=south] (FigA) at (0,0) {\includegraphics[trim=0 0 0 0,clip,width=1.24in]{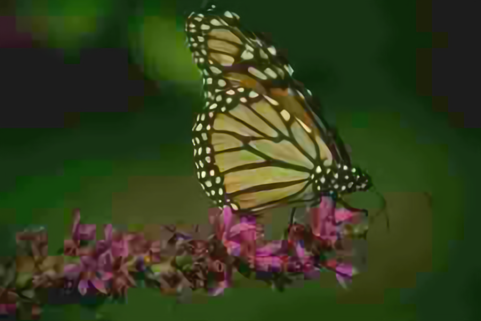}};
\spy [closeup_w_3,magnification=8] on ($(FigA)+( (+0.145, -0.035)$) 
    in node[largewindow_w,anchor=east]       at ($(FigA.north) + (-0.0402,-0.13)$);
\end{tikzpicture}
\begin{tikzpicture}[x=6cm, y=6cm, spy using outlines={every spy on node/.append style={smallwindow_w}}]
\node[anchor=south] (FigA) at (0,0)  {\includegraphics[trim=0 0 0 0,clip,width=1.24in]{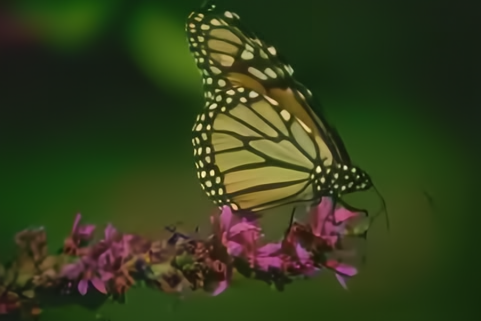}};
\spy [closeup_w_3,magnification=8] on ($(FigA)+( (+0.145, -0.035)$) 
    in node[largewindow_w,anchor=east]       at ($(FigA.north) + (-0.0402,-0.13)$);
\end{tikzpicture}
\begin{tikzpicture}[x=6cm, y=6cm, spy using outlines={every spy on node/.append style={smallwindow_w}}]
\node[anchor=south] (FigA) at (0,0)  {\includegraphics[trim=0 0 0 0,clip,width=1.24in]{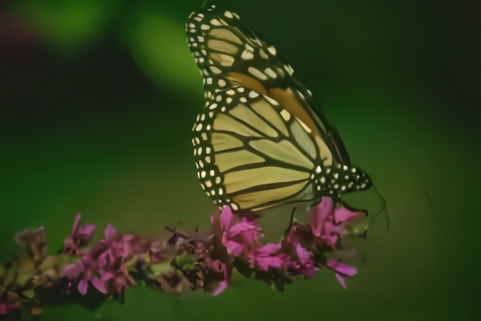}};
\spy [closeup_w_3,magnification=8] on ($(FigA)+( (+0.145, -0.035)$) 
    in node[largewindow_w,anchor=east]       at ($(FigA.north) + (-0.0402,-0.13)$);
\end{tikzpicture}
\begin{tikzpicture}[x=6cm, y=6cm, spy using outlines={every spy on node/.append style={smallwindow_w}}]
\node[anchor=south] (FigA) at (0,0) {\includegraphics[trim=0 0 0 0,clip,width=1.24in]{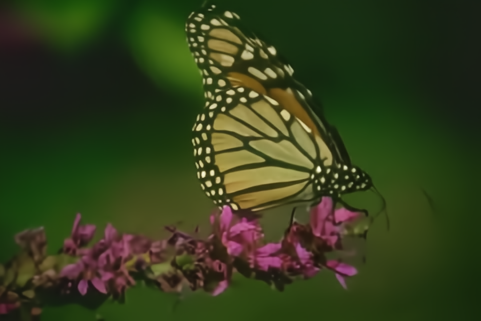}};
\spy [closeup_w_3,magnification=8] on ($(FigA)+( (+0.145, -0.035)$) 
    in node[largewindow_w,anchor=east]       at ($(FigA.north) + (-0.0402,-0.13)$);
\end{tikzpicture}
\begin{tikzpicture}[x=6cm, y=6cm, spy using outlines={every spy on node/.append style={smallwindow_w}}]
\node[anchor=south] (FigA) at (0,0)  {\includegraphics[trim=0 0 0 0,clip,width=1.24in]{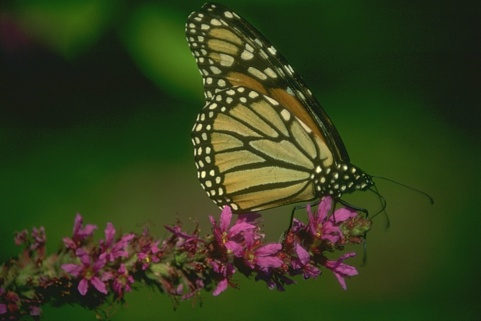}};
\spy [closeup_w_3,magnification=8] on ($(FigA)+( (+0.145, -0.035)$) 
    in node[largewindow_w,anchor=east]       at ($(FigA.north) + (-0.0402,-0.13)$);
\end{tikzpicture}
\raisebox{0.2in}{\rotatebox{90}{LIVE1 \cite{Live1}}}
\stackunder[2pt]{
\begin{tikzpicture}[x=6cm, y=6cm, spy using outlines={every spy on node/.append style={smallwindow_w}}]
\node[anchor=south] (FigA) at (0,0) {\includegraphics[trim=200 100 250 230,clip,width=1.24in]{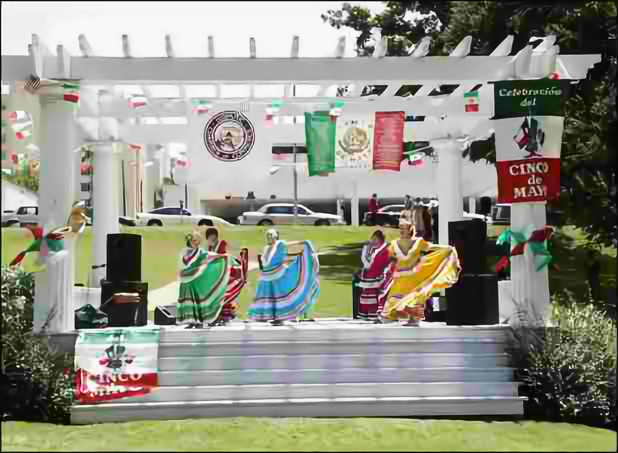}};
\spy [closeup_w_3,magnification=2] on ($(FigA)+( -0.02, 0.02)$) 
    in node[largewindow_w,anchor=east]       at ($(FigA.north) + (0.262,-0.290)$);
\end{tikzpicture}
}{DnCNN \cite{dncnn}}
\hspace{-2.2mm}
\stackunder[2pt]{
\begin{tikzpicture}[x=6cm, y=6cm, spy using outlines={every spy on node/.append style={smallwindow_w}}]
\node[anchor=south] (FigA) at (0,0) {\includegraphics[trim=200 100 250 230,clip,width=1.24in]{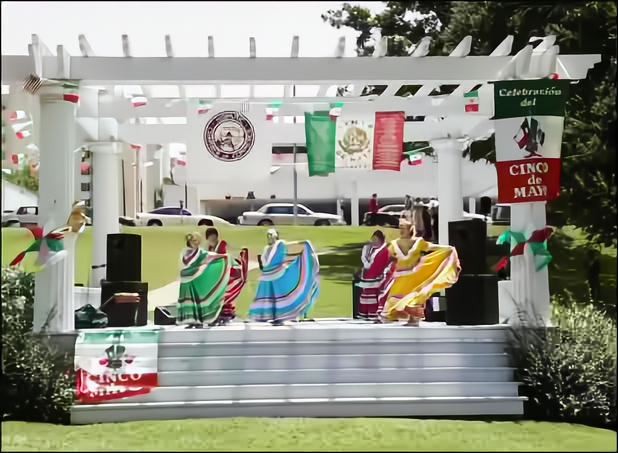}};
\spy [closeup_w_3,magnification=2] on ($(FigA)+( -0.02, 0.02)$) 
    in node[largewindow_w,anchor=east]       at ($(FigA.north) + (0.262,-0.290)$);
\end{tikzpicture}
}{QGAC \cite{qgac}}
\hspace{-2mm}
\stackunder[2pt]{
\begin{tikzpicture}[x=6cm, y=6cm, spy using outlines={every spy on node/.append style={smallwindow_w}}]
\node[anchor=south] (FigA) at (0,0) {\includegraphics[trim=200 100 250 230,clip,width=1.24in]{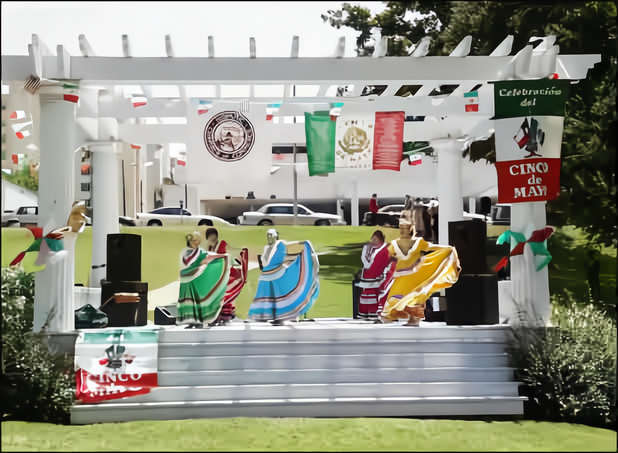}};
\spy [closeup_w_3,magnification=2] on ($(FigA)+( -0.02, 0.02)$) 
    in node[largewindow_w,anchor=east]       at ($(FigA.north) + (0.262,-0.290)$);
\end{tikzpicture}
}{FBCNN \cite{fbcnn}}
\hspace{-2mm}
\stackunder[2pt]{
\begin{tikzpicture}[x=6cm, y=6cm, spy using outlines={every spy on node/.append style={smallwindow_w}}]
\node[anchor=south] (FigA) at (0,0) {\includegraphics[trim=200 100 250 230,clip,width=1.24in]{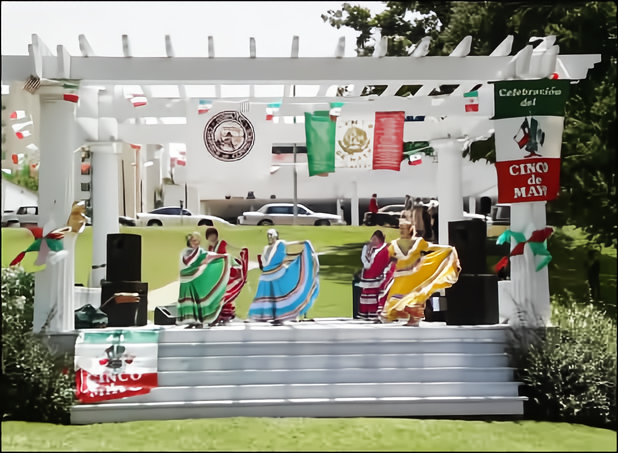}};
\spy [closeup_w_3,magnification=2] on ($(FigA)+( -0.02, 0.02)$) 
    in node[largewindow_w,anchor=east]       at ($(FigA.north) + (0.262,-0.290)$);
\end{tikzpicture}
}{\textbf{JDEC (\textit{ours})}}
\hspace{-2mm}
\stackunder[2pt]{
\begin{tikzpicture}[x=6cm, y=6cm, spy using outlines={every spy on node/.append style={smallwindow_w}}]
\node[anchor=south] (FigA) at (0,0) {\includegraphics[trim=200 100 250 230,clip,width=1.24in]{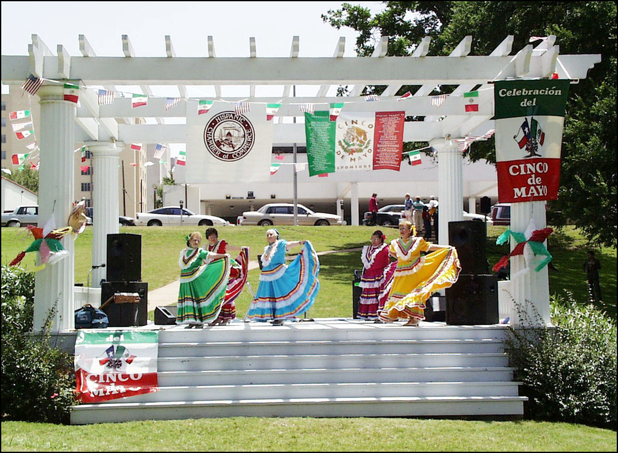}};
\spy [closeup_w_3,magnification=2] on ($(FigA)+( -0.02, 0.02)$) 
    in node[largewindow_w,anchor=east]       at ($(FigA.north) + (0.262,-0.290)$);
\end{tikzpicture}
}{{GT}}
\vspace*{-6pt}
\caption{ Qualitative comparison in color JPEG artifact removal ($q =10$). }
\vspace*{-12pt}
\label{fig:Qual_jpeggone}
\end{figure*}

\begin{figure}[t]
\footnotesize
\centering

\begin{tabular}{@{} c c  @{}}
  \begin{tabular}{@{} c @{}}
    \stackunder[2pt]{
    \begin{tikzpicture}[x=6cm, y=6cm, spy using outlines={every spy on node/.append style={smallwindow_w}}]
\node[anchor=south] (FigA) at (0,0){\includegraphics[trim={175 198 255 80},clip,width=1.2in]{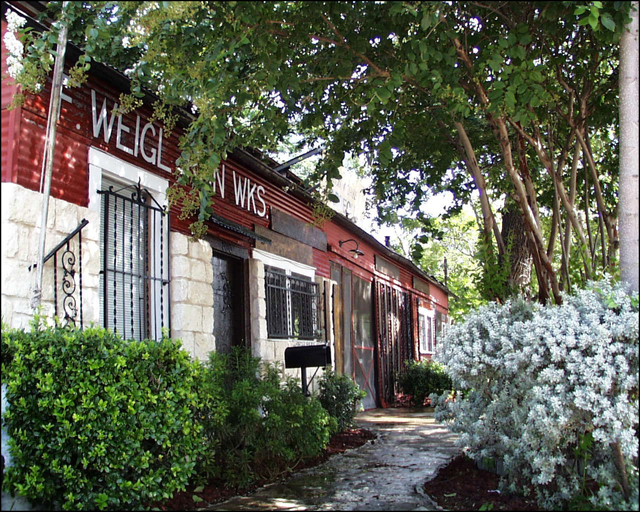}};
\spy [closeup_w_5,magnification=1.7] on ($(FigA)+( -0.068, 0.01)$) 
    in node[largewindow_w,anchor=east]       at ($(FigA.north) + (0.257,-0.455)$); 
\vspace{-10pt}
\end{tikzpicture}
}{ RGB Image}
  \end{tabular}
\hspace{-7pt}
  \begin{tabular}{@{} c c c @{}}
    \stackunder[2pt]{
        \begin{tikzpicture}
            \node[anchor=south west, inner sep=0pt] {\includegraphics[trim={220 288 355 160},clip,height=0.6in]{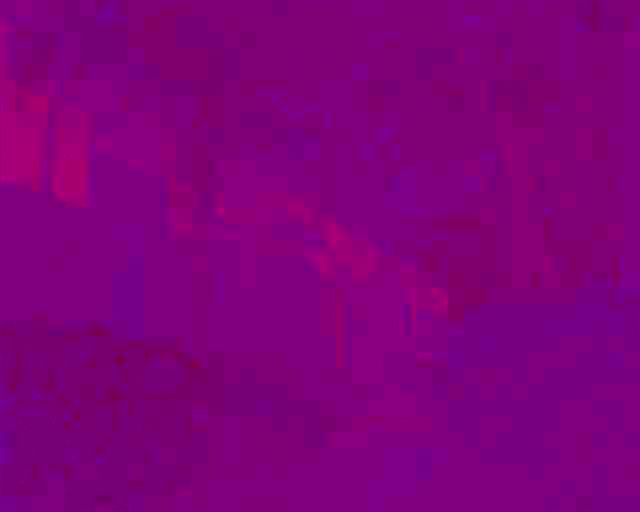}};
        \end{tikzpicture}}
    {JPEG}
    \hspace{-15pt}
 &
    \stackunder[2pt]{
        \begin{tikzpicture}
            \node[anchor=south west, inner sep=0pt] {\includegraphics[trim={220 288 355 160},clip,height=0.6in]{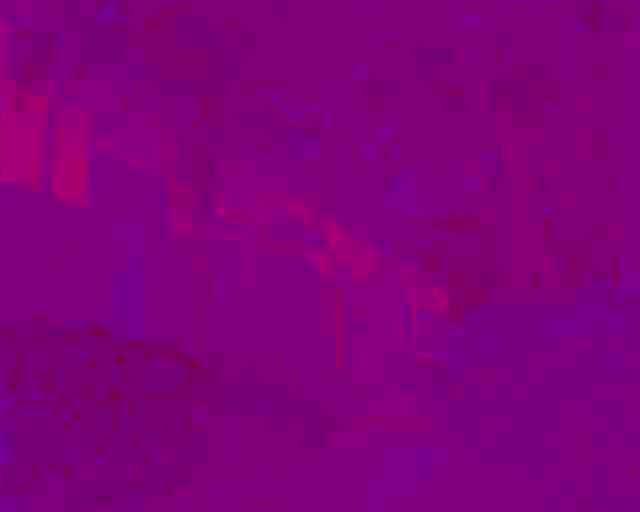}};
        \end{tikzpicture}}
    {DnCNN \cite{dncnn}}
    \hspace{-15pt}
&
        \stackunder[2pt]{
        \begin{tikzpicture}
            \node[anchor=south west, inner sep=0pt] {\includegraphics[trim={220 288 355 160},clip,height=0.6in]{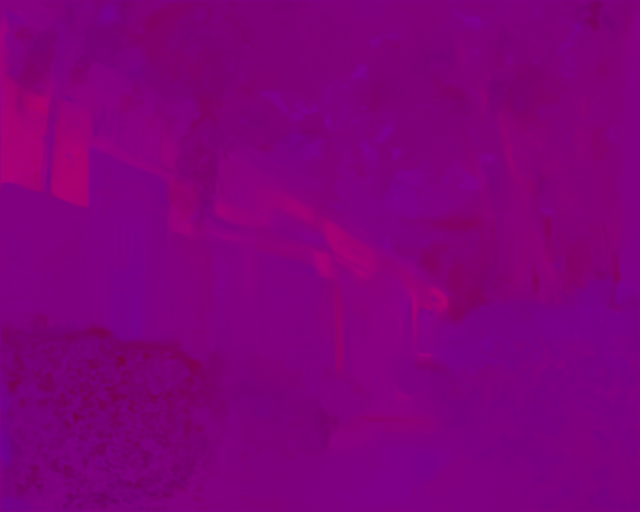}};
        \end{tikzpicture}}
    {QGAC \cite{qgac}}
\vspace{1pt}
 \\
    \stackunder[2pt]{
        \begin{tikzpicture}
            \node[anchor=south west, inner sep=0pt] {\includegraphics[trim={220 288 355 160},clip,height=0.6in]{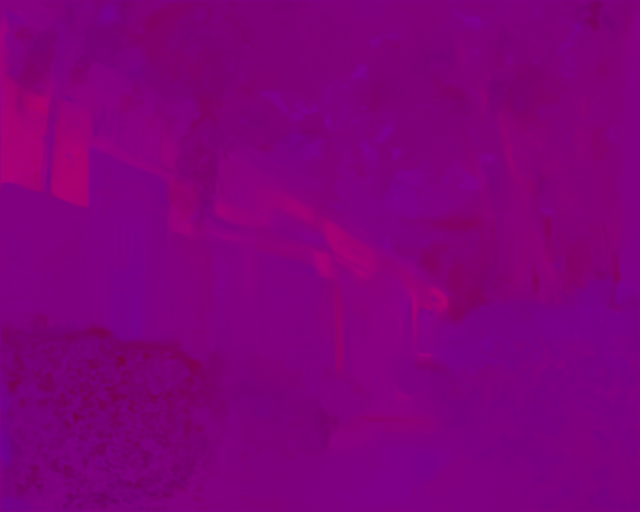}};
        \end{tikzpicture}}
    {FBCNN \cite{fbcnn}}
    \hspace{-15pt}
 &
    \stackunder[2pt]{
        \begin{tikzpicture}
            \node[anchor=south west, inner sep=0pt] {\includegraphics[trim={220 288 355 160},clip,height=0.6in]{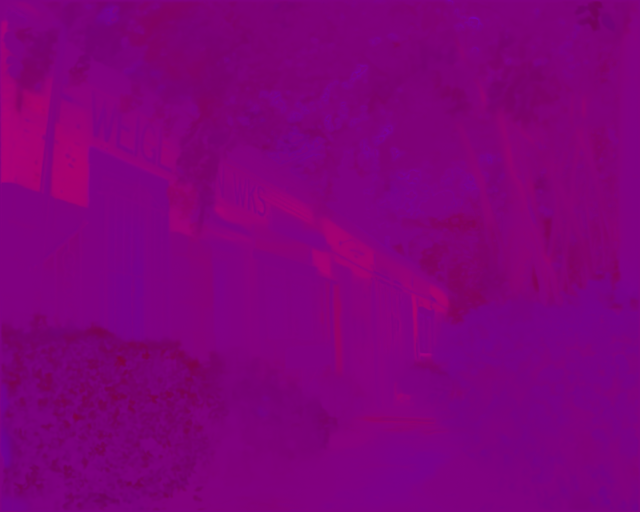}};
        \end{tikzpicture}}
    {\textbf{JDEC (\textit{ours})}}
    \hspace{-15pt}
&
    \stackunder[2pt]{
        \begin{tikzpicture}
            \node[anchor=south west, inner sep=0pt] {\includegraphics[trim={220 288 355 160},clip,height=0.6in]{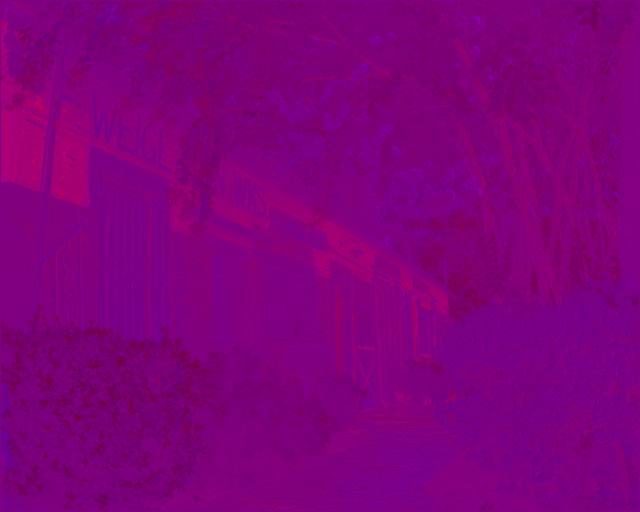}};
        \end{tikzpicture}}
    {GT ($\mathbf{I_C}$)}

  \end{tabular}

\end{tabular}
\begin{tabular}{@{} c c  @{}}
  \begin{tabular}{@{} c @{}}
    \stackunder[2pt]{
    \begin{tikzpicture}[x=6cm, y=6cm, spy using outlines={every spy on node/.append style={smallwindow_w}}]
\node[anchor=south] (FigA) at (0,0){\includegraphics[trim={162 40 174 0},clip,width=1.2in]{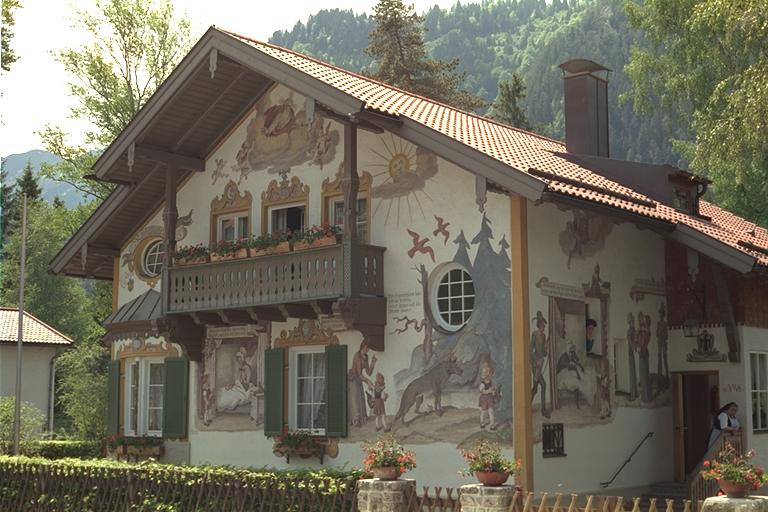}};
\spy [closeup_w_5,magnification=6] on ($(FigA)+( 0.008, 0.16)$) 
    in node[largewindow_w,anchor=east]       at ($(FigA.north) + (0.257,-0.445)$);
\end{tikzpicture}
}{RGB Image}
  \end{tabular}
\hspace{-7pt}

  \begin{tabular}{@{} c c c @{}}
    \stackunder[2pt]{
        \begin{tikzpicture}
            \node[anchor=south west, inner sep=0pt] {\includegraphics[trim={362 400 374 80},clip,height=0.6in]{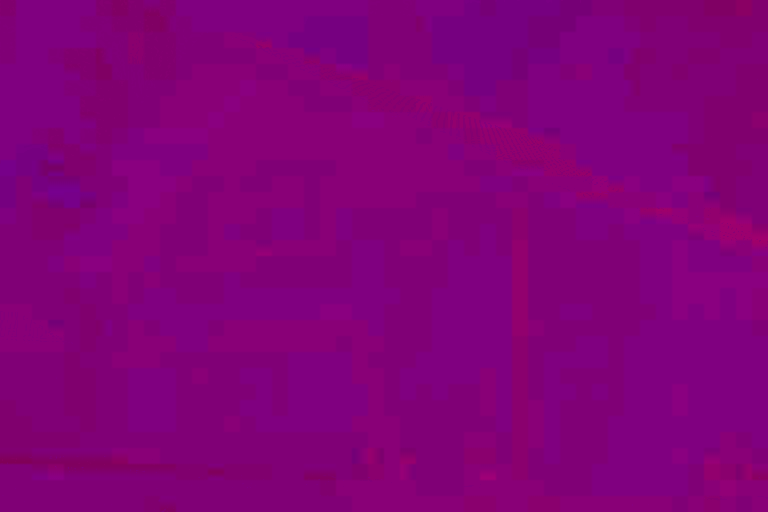}};
        \end{tikzpicture}}
    {JPEG}
    \hspace{-15pt}
 &
    \stackunder[2pt]{
        \begin{tikzpicture}
            \node[anchor=south west, inner sep=0pt] {\includegraphics[trim={362 400 374 80},clip,height=0.6in]{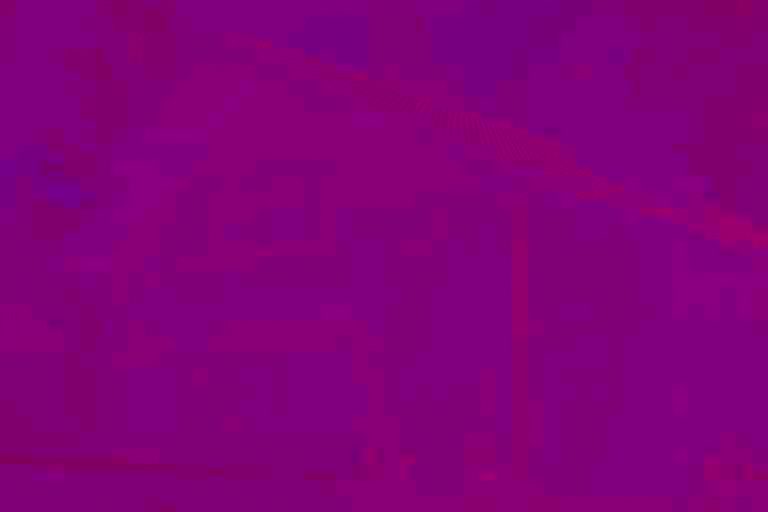}};
        \end{tikzpicture}}
    {DnCNN \cite{dncnn}}
    \hspace{-15pt}
&
        \stackunder[2pt]{
        \begin{tikzpicture}
            \node[anchor=south west, inner sep=0pt] {\includegraphics[trim={362 400 374 80},clip,height=0.6in]{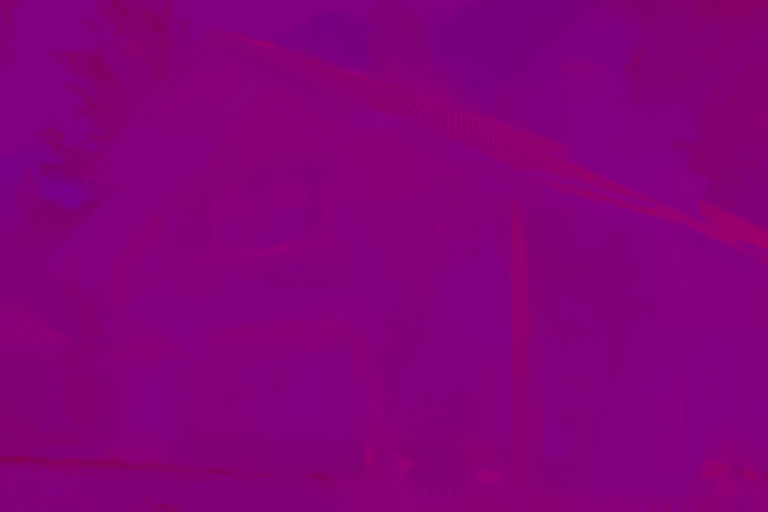}};
        \end{tikzpicture}}
    {QGAC \cite{qgac}}
\vspace{1pt}
 \\
    \stackunder[2pt]{
        \begin{tikzpicture}
            \node[anchor=south west, inner sep=0pt] {\includegraphics[trim={362 400 374 80},clip,height=0.6in]{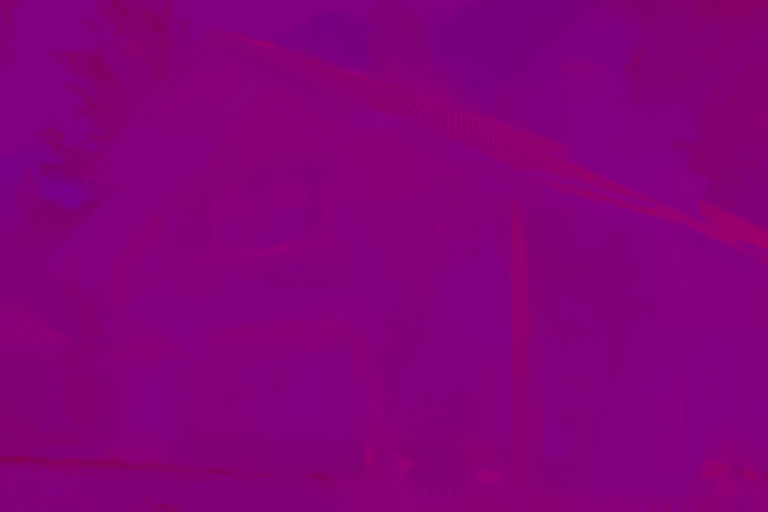}};
        \end{tikzpicture}}
    {FBCNN \cite{fbcnn}}
    \hspace{-15pt}
 &
    \stackunder[2pt]{
        \begin{tikzpicture}
            \node[anchor=south west, inner sep=0pt] {\includegraphics[trim={362 400 374 80},clip,height=0.6in]{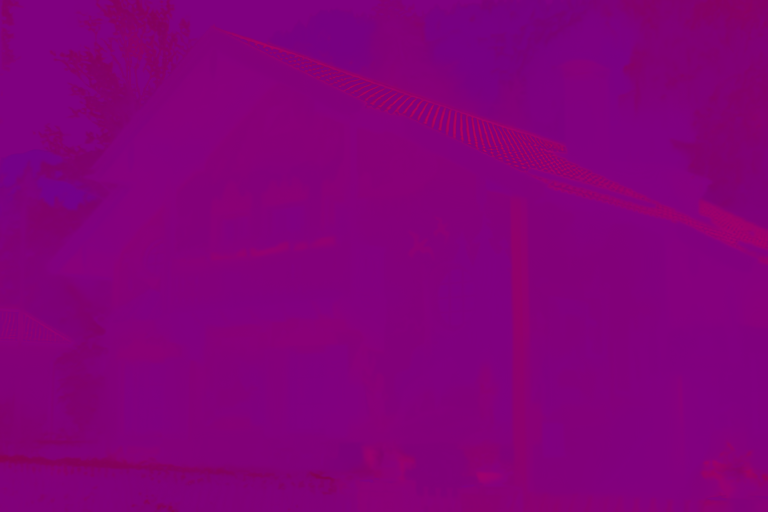}};
        \end{tikzpicture}}
    {\textbf{JDEC (\textit{ours})}}
    \hspace{-15pt}
&
    \stackunder[2pt]{
        \begin{tikzpicture}
            \node[anchor=south west, inner sep=0pt] {\includegraphics[trim={362 400 374 80},clip,height=0.6in]{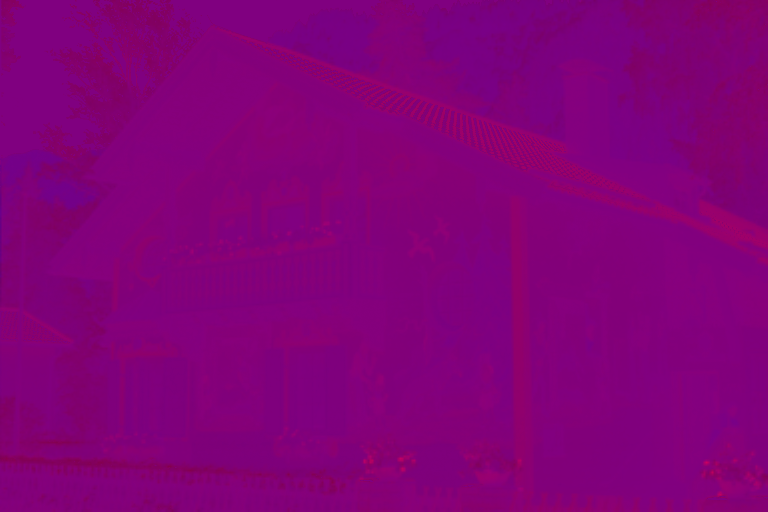}};
        \end{tikzpicture}}
    {GT ($\mathbf{I_C}$)}

  \end{tabular}

\end{tabular}

\hspace{0.3mm}
\vspace*{-8pt}
\caption{ Qualitative comparison in chroma components $\mathbf{I_c}$ of images ($q =10$). }
\vspace*{-23pt}
\label{fig:Qual_chroma}
\end{figure}

\subsection{Training}

\noindent \textbf{Dataset} Following the previous work \cite{qgac,fbcnn}, we use DIV2K and Flickr2K \cite{DIV2kdataset}. Each dataset contains 800 and 2650 images, respectively. For generating synthetic JPEG compression, we use the OpenCV standard \cite{opencv_library}. We compress images using randomly sampled quality factors with steps of 10 in the range [10,100]. We directly extract quantization maps $\mathbf{Q}$ and coefficients of spectra $\mathbf{\Tilde{C}}$ from JPEG files and construct spectra $\mathbf{\Tilde{X}}$, following the \cref{eq:quantize_spectra}. Since the dynamic range of spectra depends on frequency, we should normalize spectra in a range of $[-1,1]$. The quantization maps are normalized with the same normalization function. The ground truth (GT) images are prepared with a range of $[-0.5,0.5]$ because the JPEG encoder subtracts the midpoint of the image range (=128).

\noindent \textbf{Implementation Detail} 
We use $112\times112$ patches as inputs to our network. This size is chosen because it is the least common multiple of the minimum unit size of color JPEG ($16\times16$) and the window size of our Swin architecture \cite{Liu_2022_CVPR_swin2} ($7\times7$). The network is trained for 1000 epochs with batch size 16. We optimize our network by Adam \cite{DBLP:journals/corr/KingmaB14}. The learning rate is initialized as 1e-4 and decayed by factor 0.5 at $[200,400,600,800]$.

\begin{table}[ht]
\footnotesize
\vspace{-5pt}
\centering
\setlength{\tabcolsep}{1.2pt}
\begin{tabular}{c
|>{\centering\arraybackslash}p{1.4cm}
|>{\centering\arraybackslash}p{1.4cm}
|>{\centering\arraybackslash}p{1.4cm}
|>{\centering\arraybackslash}p{1.4cm}
}
&\multicolumn{4}{c}{{Luma$(\mathbf{I_Y})|$Chroma$\mathbf{(I_C)}$}}\\
\hline
Method & $q=10$&$q=20$&$q=30$&$q=40$ \\
\hline\hline
{JPEG} & 34.39$|$35.77 &  37.32$|$38.90& 39.85$|$40.72  &39.82$|$39.59\\
{DnCNN \cite{dncnn}} & 35.30$|$35.85 & 37.60$|$38.01& 38.78$|$38.96  &39.45$|$39.51\\
{QGAC \cite{qgac}} & \textcolor{blue}{37.28}$|$38.18 & \textcolor{blue}{39.75}$|$39.94&\textcolor{blue}{41.00}$|$40.69 & \textcolor{blue}{41.73}$|$41.09  \\
{FBCNN \cite{fbcnn}}  & \textcolor{black}{37.12}$|$\textcolor{blue}{38.36} & \textcolor{black}{39.71}$|$\textcolor{blue}{40.21}& \textcolor{black}{40.97}$|$\textcolor{blue}{41.04} &  \textcolor{black}{41.81}$|$\textcolor{blue}{41.50}\\
{\textbf{JDEC} (\textit{ours})} & \textcolor{red}{37.32}$|$\textcolor{red}{38.90} & \textcolor{red}{39.85}$|$\textcolor{red}{40.72}  & \textcolor{red}{41.11}$|$\textcolor{red}{41.52} & \textcolor{red}{41.92}$|$\textcolor{red}{41.96}\\
\end{tabular}
\vspace*{-6pt}
\caption{{Quantitative comparisons of each components in ICB \cite{ICBdataset} datasets. (PSNR(dB))}}\label{tab:chroma-comparison}
\vspace{-20pt}
\end{table}

\subsection{Evaluation}
\noindent \textbf{Quantitative Result} \quad For evaluation, we use LIVE-1 \cite{Live1}, testset of BSDS500 \cite{B500dataset} and ICB \cite{ICBdataset} dataset. In the aspect of the JPEG decoder, we present the rate-distortion curve to illustrate the trade-off between bits-per-pixel (bpp) and peak signal-to-noise ratio (PSNR) where quality factors in a range of $[10,100]$. We observed that BSDS500 \cite{B500dataset} is saved as JPEG with a quality factor of 95. Therefore, the reported BSDS500 data is within a quality factor of 90. We compare our JDEC against existing compression artifact removal models: DnCNN \cite{dncnn}, QGAC \cite{qgac}, and FBCNN \cite{fbcnn} in \cref{fig:RDcurve-qual}. The selected models cover a relatively wide range of quality factors with a single network. We evaluate DnCNN \cite{dncnn} following the suggested method in QGAC \cite{qgac}, with channels being processed independently. Despite QGAC \cite{qgac} having a training range of $[10,100]$, it experiences a drop in performance in the range of $\left(90,100\right]$ across all datasets. FBCNN \cite{fbcnn} also exhibits a performance drop in the range of $[95,100]$ when evaluated on the LIVE-1 \cite{Live1} dataset. In comparison, JDEC outperforms all other methods, regardless of the quality factor or dataset.

Regarding JPEG artifact removal, we report PSNR, structural similarity index (SSIM), and PSNR-B for estimating de-blocking in \cref{tab:Quan_testvalid}. We include DMCNN \cite{dmcnn}, IDCN \cite{idcn}, and transformer-based Swin2SR \cite{conde2022swin2sr} as additional comparative groups since they cover a range of quality factors. Note that the Swin2SR has trained on a limited range of quality factors in a range of $[10,40]$ with additional datasets, including the train and test dataset of BSDS500 \cite{B500dataset} and Waterloo \cite{ma2017waterloo}. We partitioned the data presented in \cref{tab:Quan_testvalid} to distinguish between networks operating within limited and expansive ranges. Our JDEC shows remarkable performance compared to other methods. The maximum PSNR interval is 0.37dB on ICB for $q=10$. 

We demonstrate \cref{tab:chroma-comparison} to observe the restoration effects of two components of different sizes $\mathbf{I_Y}\in \mathbb{R}^{H\times W \times1},\mathbf{I_C}\in \mathbb{R}^{\frac{H}{2}\times\frac{W}{2}\times2}$. According to \cref{tab:chroma-comparison}, the performance difference in the chroma component $\mathbf{I_C}$ is greater than the difference of the luma component $\mathbf{I_Y}$ indicating an empirical upsampling effect.

\begin{table}[t]
\scriptsize
\vspace{-6pt}
\centering
\setlength{\tabcolsep}{1.2pt}
\begin{tabular}{c
|>{\centering\arraybackslash}p{1.5cm}
|>{\centering\arraybackslash}p{1.5cm}
|>{\centering\arraybackslash}p{1.5cm}
|>{\centering\arraybackslash}p{1.5cm}
}
{Test}&\multicolumn{4}{c}{LIVE-1 \cite{Live1}}\\
\hline
Method & $q=80$&$q=90$&$q=95^*$&$q=100$ \\
\hline\hline
\multirow{2}{*}{JPEG} & 34.23$|$33.45 &  36.86$|$36.45& 39.33$|$38.90  &\textcolor{blue}{43.07}$|$\textcolor{blue}{42.37}\\
&0.948 & 0.967 &0.979 &\textcolor{blue}{0.993} \\
\multirow{2}{*}{DNCNN \cite{dncnn}}   &35.01$|$34.69 & 37.29$|$36.97&  39.20$|$38.79 & 41.15$|$40.59\\
 &  0.954&  0.970&\textcolor{blue}{0.980} & 0.987\\
\multirow{2}{*}{QGAC \cite{qgac}} & 35.75$|$35.19 & 37.75$|$37.20&37.50$|$ 37.01 & 38.97$|$38.56  \\
&0.960 &0.973&\textcolor{black}{0.974}  & 0.979\\
\multirow{2}{*}{FBCNN \cite{fbcnn}}  & \textcolor{blue}{36.02}$|$\textcolor{blue}{35.41} & \textcolor{blue}{38.25}$|$\textcolor{blue}{37.68}& \textcolor{blue}{40.23}$|$\textcolor{blue}{39.65} &  42.23$|$41.52\\
& \textcolor{blue}{0.961}& \textcolor{blue}{0.974}&\textcolor{red}{0.983}& 0.990\\
\multirow{2}{*}{\textbf{JDEC} (\textit{ours})} & \textcolor{red}{36.31}$|$ \textcolor{red}{35.73} & \textcolor{red}{38.72}$|$\textcolor{red}{38.17}  & \textcolor{red}{40.41}$|$\textcolor{red}{39.90} & \textcolor{red}{45.14}$|$\textcolor{red}{44.20}\\
 & \textcolor{red}{0.963}& \textcolor{red}{0.976}&\textcolor{red}{0.983}&\textcolor{red}{0.995} \\
\hline
\hline
{Test}&\multicolumn{4}{c}{ICB \cite{ICBdataset}}\\
\hline
Method & $q=80$&$q=90$&$q=95^*$&$q=100$ \\
\hline\hline
\multirow{2}{*}{JPEG}  & {36.34$|$35.82} & 37.72$|$ 37.40& {39.17$|$39.01} &41.31$|$41.28\\
 &0.891 &  0.912&0.934&\textcolor{blue}{0.955}\\
\multirow{2}{*}{DNCNN \cite{dncnn}}  & 35.57$|$35.44 & 36.75$|$36.64 & 37.99$|$37.92 & 39.73$|$ 39.69\\
& 0.844&0.868& 0.891 & 0.915 \\
\multirow{2}{*}{QGAC \cite{qgac}} & 37.58$|$37.47 &38.34$|$38.21& 36.84$|$36.68 &37.55$|$37.48  \\
 & \textcolor{blue}{0.902}&0.919& \textcolor{black}{0.912} & 0.926\\
\multirow{2}{*}{FBCNN \cite{fbcnn}} & \textcolor{blue}{38.03}$|$\textcolor{blue}{37.91} & \textcolor{blue}{39.17}$|$\textcolor{blue}{39.03} & \textcolor{blue}{40.36}$|$\textcolor{blue}{40.22} & \textcolor{blue}{41.61}$|$\textcolor{blue}{41.52}\\
& \textcolor{blue}{0.902}& \textcolor{blue}{0.920} &\textcolor{blue}{0.938}&  \textcolor{black}{0.951}\\
\multirow{2}{*}{\textbf{JDEC} (\textit{ours})}  & \textcolor{red}{38.43}$|$\textcolor{red}{38.29} & \textcolor{red}{39.58}$|$\textcolor{red}{39.41} & \textcolor{red}{40.77}$|$\textcolor{red}{40.63}& \textcolor{red}{43.61}$|$\textcolor{red}{43.52}\\
 & \textcolor{red}{0.906}& \textcolor{red}{0.924} &\textcolor{red}{0.943}&\textcolor{red}{0.968} \\
\end{tabular}
\vspace*{-6pt}
\caption{Quantitative comparisons of high-quality images in LIVE-1 \cite{Live1} and ICB \cite{ICBdataset} datasets (PSNR$|$PSNR-B(dB)) (top), SSIM (bottom). *: Quality factor 95 is a generally used default quality factor in the JPEG encoder.}\label{tab:high-quality-comparison}
\vspace{-10pt}
\end{table}


In \cref{tab:high-quality-comparison}, we show the comparison of the high-quality image decoding. The $^*$ mark indicates the commonly used default quality factor of the JPEG, including OpenCV \cite{opencv_library}. As a practical decoder for JPEG, only our {JDEC} decodes the best images among other baselines, including the conventional JPEG decoder. 

\def\checkmark{\tikz\fill[scale=0.4](0,.35) -- (.25,0) -- (1,.7) -- (.25,.15) -- cycle;}
\begin{table}[t]
\vspace{-3pt}
\centering
\setlength{\tabcolsep}{1.2pt}
\footnotesize
\begin{tabular}{c
|>{\centering\arraybackslash}p{1cm}
|>{\centering\arraybackslash}p{1cm}
|>{\centering\arraybackslash}p{1.33cm}
|>{\centering\arraybackslash}p{1.33cm}
|>{\centering\arraybackslash}p{1.33cm}
|>{\centering\arraybackslash}p{1.33cm}
}
\multirow{2}{*}{ID}&\multicolumn{2}{c|}{Method}&\multicolumn{4}{c}{Quality Factor $q$}\\
\cline{2-7}
                      &$g_\phi$-(a) & $\mathbf{T}_\psi$ &10& 20&30&40 \\
\hline\hline
$0^*$&\checkmark&\checkmark & \textbf{27.95$|$27.71} & \textbf{30.26$|$29.87} &  \textbf{31.59$|$31.12} &\textbf{32.50$|$31.98}\\
\hline
1&\checkmark&  \xmark  & 27.76$|$27.51 & 30.04$|$29.62 & 31.35$|$30.84 & 32.25$|$31.68\\
2& \xmark &\checkmark& 27.69$|$27.43 & 29.95$|$29.53 &31.25$|$30.71 & 32.14$|$31.54  \\
3& \xmark & \xmark & 26.90$|$26.61 & 28.37$|$28.06 & 28.73$|$28.36 & 28.96$|$28.57\\
\hline
4& \checkmark & \cref{eq:lte_dft} & 27.88$|$27.64 & 30.21$|$29.83 & 31.54$|$31.07 & 32.45$|$31.93\\
\end{tabular}
\vspace*{-6pt}
\caption{\textbf{Quantitative ablation study} of JDEC on LIVE-1 \cite{Live1} (PSNR$|$PSNR-B (dB)). $*$: ID-0 is the proposed method JDEC. The definition of each ID number is shown in \cref{sec:abl}.}\label{tab:Quan_abl_tab}
\vspace{-20pt}
\end{table}


\noindent \textbf{Qualitative Result} We show color JPEG artifact removal task in \cref{fig:Qual_jpeggone}. There are two main distortions of JPEG compression: 1) lack of High-frequency components and 2) color differences. We demonstrate the effect of our JDEC in addressing the distortions in high frequencies in the first row of \cref{fig:Qual_jpeggone}. While other methods suffer from aliasing, our JDEC successfully recovers the details of the butterfly's antennae. In the second row of \cref{fig:Qual_jpeggone}, our JDEC relieves color distortion derived from JPEG compression.

In \cref{fig:Qual_chroma}, we sort out the chroma components from each image to demonstrate the effect of our JDEC in relieving color distortions. When observing the chroma components using other methods, it is noticeable that the chroma components remain significantly distorted. However, our JDEC restores them closer to the original. It demonstrates that JDEC robustly restores color components subjected to quantization and downsampling, effectively mitigating distortion.

\begin{figure}[t]
    \vspace{-5pt}
    \centering
    \includegraphics[trim=0 0 20 0,clip,width = 3.2in]{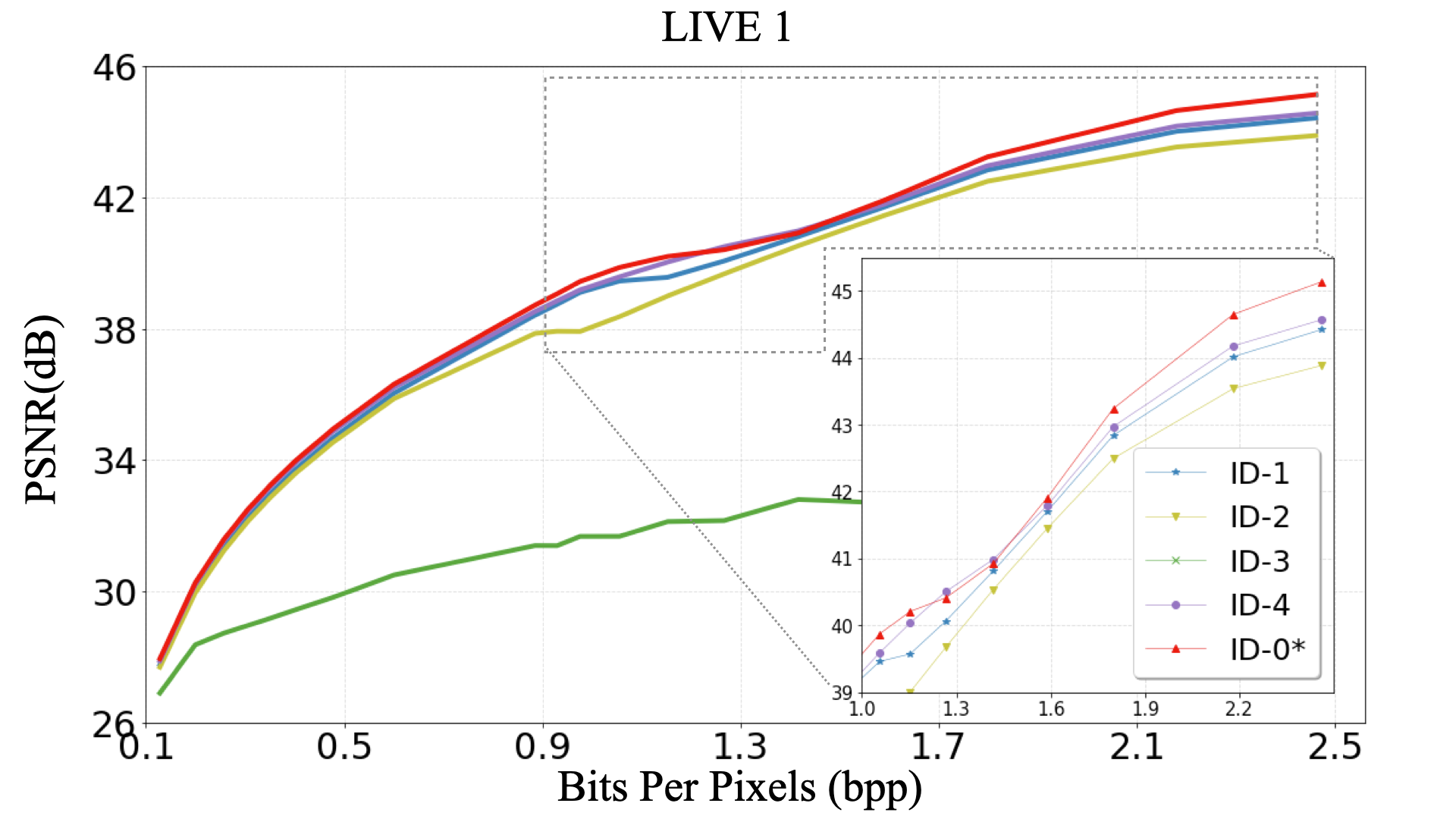}
    \vspace{-8pt}
    \caption{\textbf{Quantitative ablation study} of JDEC on LIVE-1 \cite{Live1} (RD-curve), against ablation models. Our proposed JDEC achieves higher PSNR than any other models in most of the $q$ values.}
    \label{fig:ablationrd}
    \vspace{-20pt}
\end{figure}

\subsection{Ablation Study}\label{sec:abl}

\noindent \textbf{Network Components }We conducted ablation studies for the main components of our proposed JDEC. The proposed method, CCF $\mathbf{T}_\psi$ contains a frequency estimator which makes JDEC learn enhanced spectra. To support this, we train JDEC without a frequency estimator, directly forwarding concatenated coordinates (ID-1). We use additional 3$\times$3 convolutional layers to have a comparable number of parameters.
The sub-block conversion is the main element of encoder $\mathbf{E}_\varphi$. The spatial area gains a degree of freedom by using the sub-block conversion of the DCT matrix. We conduct the ablation study of sub-block conversion by embedding inputs directly (ID-2 of \cref{tab:Quan_abl_tab}). The drop in performance is severe when both components are missing (ID-3 of \cref{tab:Quan_abl_tab}). We also observed that  ID-3, training without both the group embedding and CCF, leads to a significant performance drop as shown in \cref{fig:ablationrd}.

\noindent \textbf{Fourier Features} Comparing to the existing sinusoidal representation, the formulation of \cite{lee2021local} will be compatible for our CCF. The modified Fourier feature is as follows :
\begin{equation}
\mathbf{{C}}
\odot
\begin{bmatrix}
\cos(\pi ({{\mathbf{F}\cdot\delta}}+h_q({\mathbf{Q}}))\\
\sin(\pi ({{\mathbf{F}\cdot\delta}}+h_q({\mathbf{Q}}))
\end{bmatrix} .
\vspace{-6pt}\label{eq:lte_dft}
\end{equation}
We label the model using \cref{eq:lte_dft} instead of $\mathbf{T}_\psi$ as ID-4.
The rate-distortion curve of all ablation models is illustrated in \cref{fig:ablationrd}. As shown in \cref{fig:ablationrd}, the maximum gain of our CCF is 0.58dB against ID-4, where the quality factor is 100. \cref{eq:lte_dft} is considered as using additional terms than ID-0 (JDEC) by trigonometric sum. However, it has led to performance degradation as shown in \cref{tab:Quan_abl_tab}. 

\subsection{Continuous Cosine Spectrum}\label{sec:demo_CCF}

\begin{figure}[t]
  \scriptsize
    \centering
    \begin{subfigure}[t]{0.3\textwidth}
    \hspace{-40pt}
         \stackunder[2pt]{\hspace{-8pt} JPEG (Input)\qquad \quad GT \qquad \quad  \textbf{JDEC}\textit{(ours)}   \quad\quad $\mathbf{\widehat{X}(F_h,F_w)}$ in \cref{fig:graphical} }
         {\includegraphics[trim=119 350 0 58,clip,width = 3.3in]{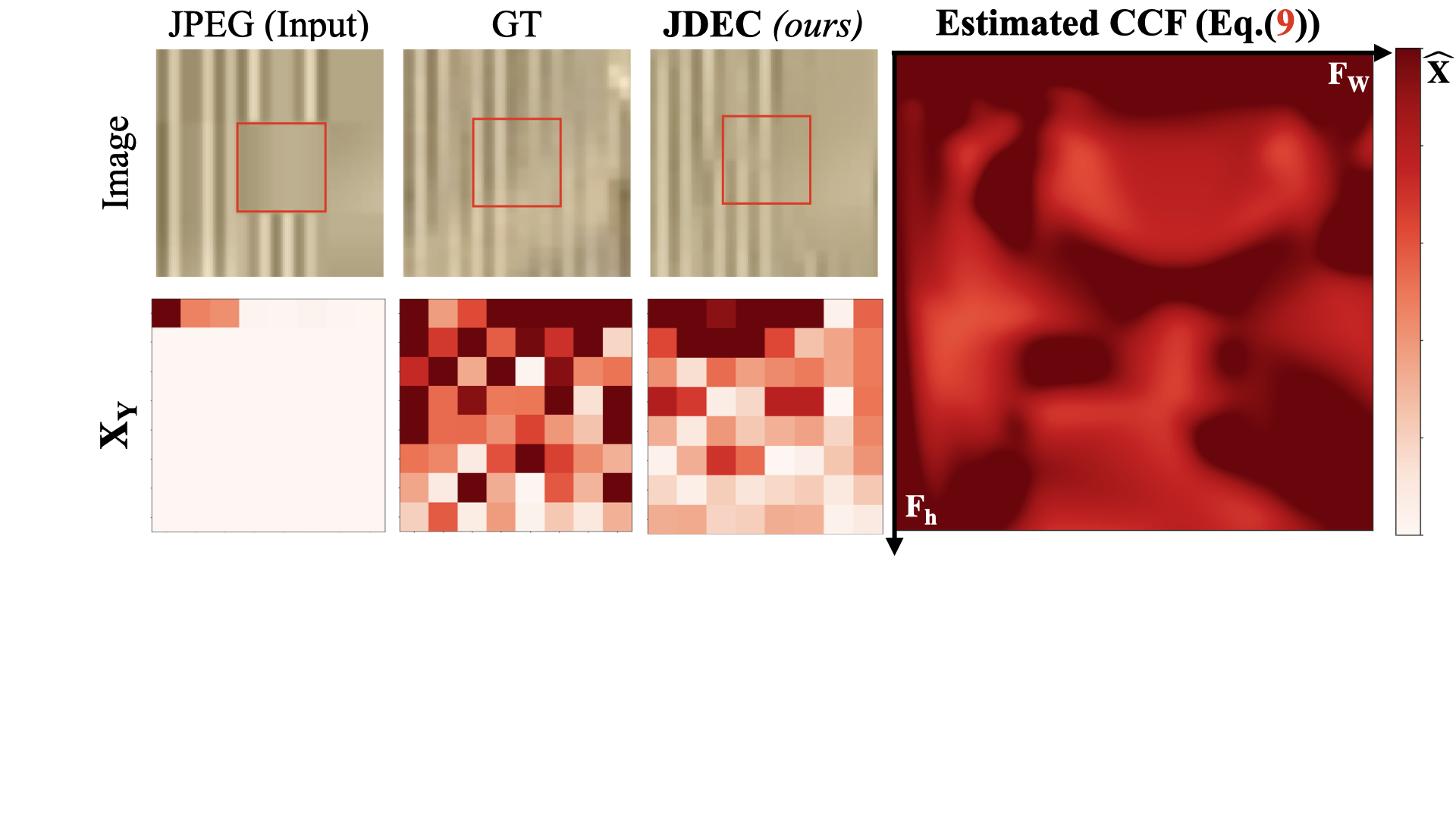}}
        \caption{High-frequency}
        \label{fig:spectra-high}
        \vspace{-2pt}
    \end{subfigure}
    \begin{subfigure}[t]{0.3\textwidth}
    \hspace{-40pt}
         \stackunder[2pt]{\hspace{-8pt} JPEG (Input)\qquad \quad GT \qquad \quad  \textbf{JDEC}\textit{(ours)}   \quad\quad $\mathbf{\widehat{X}(F_h,F_w)}$ in \cref{fig:graphical} }
         {\includegraphics[trim=119 350 0 58,clip,width = 3.3in]{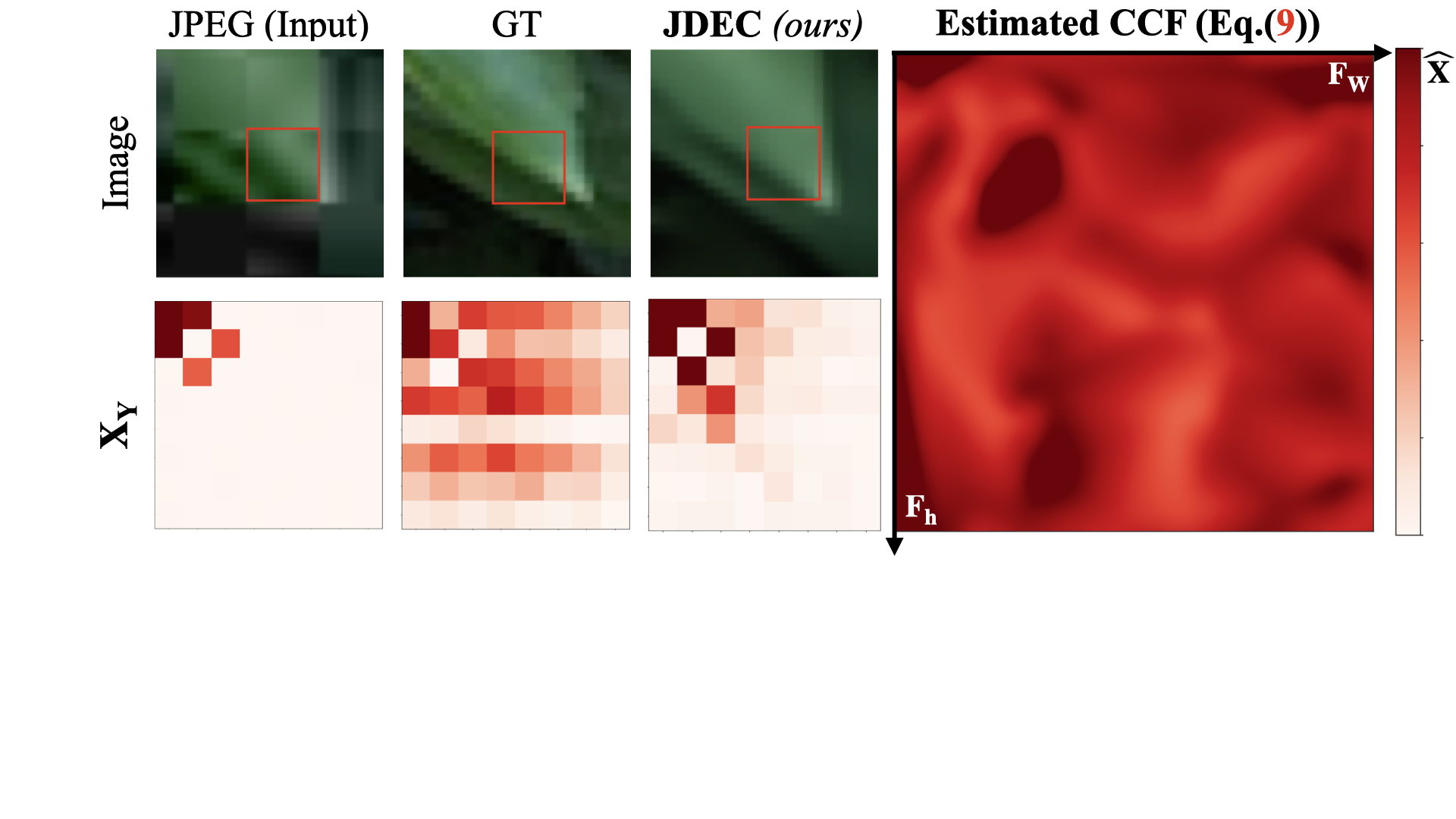}}
        \caption{Low-frequency}
        \label{fig:spectra-low}
        \vspace{-10pt}
    \end{subfigure}
    \caption{\textbf{Comparison of the estimated spectra of the Continuous Cosine Formulation (CCF).} The quality factor of input is 10. The estimated CCF spectrum follows the spectrum of ground-truth images despite severe distortion.}
    \label{fig:spectra}
    \vspace{-19pt}
\end{figure}

In this section, we demonstrate that our CCF extracts dominant frequencies and amplitudes from highly compressed JPEG spectra. The ranges of input images (8$\times$8) are highlighted with red boxes in each image. For visualization, we observe components of CCF, including estimated frequencies $\mathbf{F_h, F_w}$ and amplitudes $\mathbf{\widehat{X}}$. We scatter frequencies in 2D space and assign a color to each amplitude. We quantize the frequencies to [0, 50] with steps of 1 and interpolate to continuous values. 
 In \cref{fig:spectra-high}, most of the high-frequency components have been removed. The estimated spectrum with CCF is centered on high-frequency components despite such circumstances. In the case of \cref{fig:spectra-low}, the dominant components of the spectrum are focused on relatively low-frequency. Even in this case, the extracted spectrum of CCF is concentrated in the low-frequency elements as in the ground truth.
 


\section{Discussion}

\noindent\textbf{Implicit Neural Representation} As discussed in \cref{sec:related} and \cref{eq:overall}, a JPEG encoder downsamples chroma components. Therefore, the JPEG decoder should map: $\mathbb{R}^{\frac{H}{2}\times\frac{W}{2}\times 2} \mapsto \mathbb{R}^{H\times W\times 2}$ for chroma. Our JDEC addresses this issue through CCF ($T_\psi$) by embedding $\delta$ into $1\times1$-sized features, making the proposed JDEC a function of $\delta$. Our model is able to decode high-resolution images when provided with dense coordinates that were not observed during training. We show the advanced additional upsampling results in \cref{sec:supp_inr}.

\begin{figure}[t]
\footnotesize
\centering
\stackunder[2pt]{JPEG \textbf{(bpp : 0.052)}}{\includegraphics[trim= 250 0 50 0,clip,width = 1.03in]{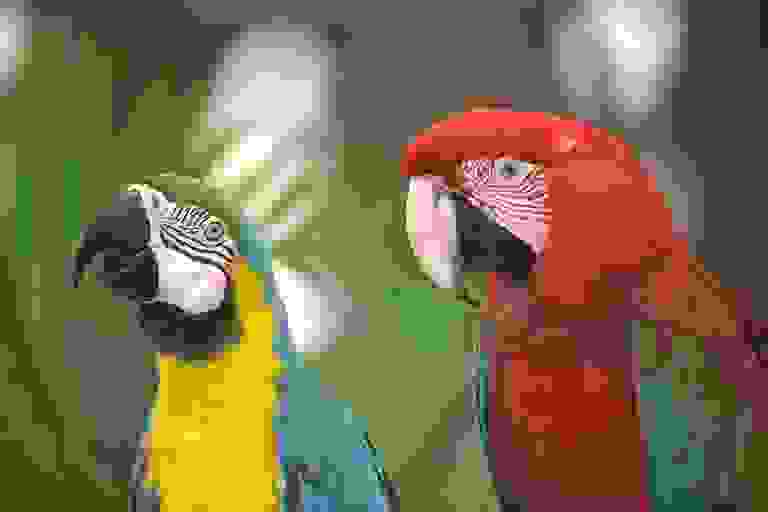}}
\hspace{1pt}
\stackunder[2pt]{FBCNN\cite{fbcnn}}{\includegraphics[trim= 250 0 50 0 ,clip,width = 1.03in]{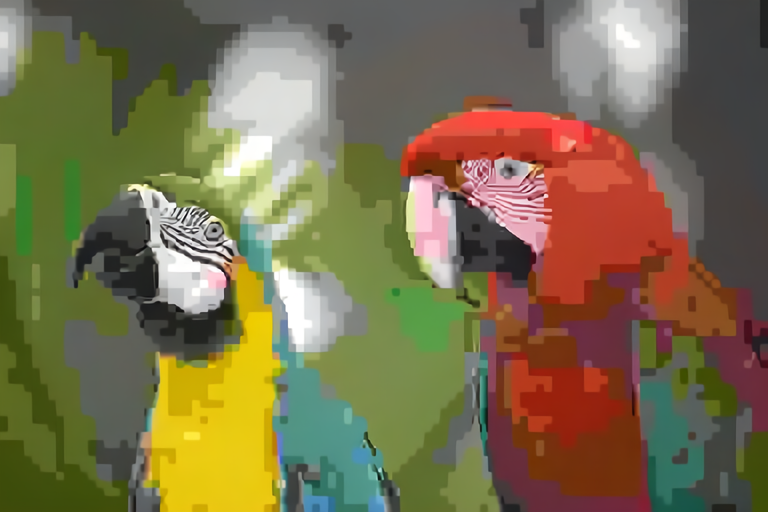}}
\hspace{1pt}
\stackunder[2pt]{JDEC+}{\includegraphics[trim= 250 0 50 0 ,clip,width = 1.03in]{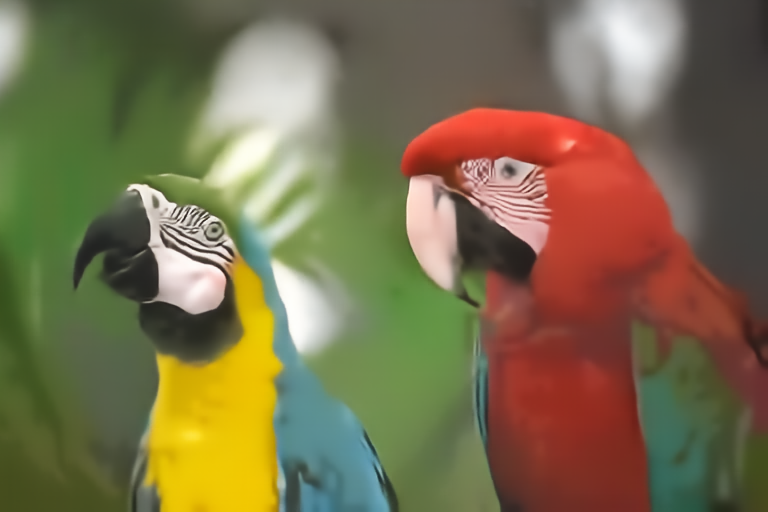}}
\vspace{1mm}
{22.62 (dB) \qquad \qquad \qquad 23.96 (dB) \qquad \qquad \qquad {\color{black}\textbf{26.70 (dB)}}}
\vspace{-4pt}
\caption{\textbf{Reconstruction of the extremely compressed image ($q=0$)} in LIVE-1 \cite{Live1} dataset.}
\vspace*{-0pt}
\label{fig:extreme}
\vspace{-4pt}
\end{figure}

\noindent\textbf{Extreme Reconstruction} We primarily propose a decoding network to generate high-quality images due to its practical applicability. Consequently, we pursued the network without explicitly considering the scenario of high compression ($q=0$). However, by incorporating all image quality factors within the range [0,100] with a step size of 10 during the learning process, we successfully developed a decoding method tailored for highly compressed images. We label the additional network as JDEC+. As shown in \cref{fig:extreme}, our JDEC+ recovers the highly compressed images better than image restoration models.

\noindent\textbf{Computation Time and Memory} In \cref{tab:model_complexity}, we report computational resources including the number of parameters, memory consumption, floating-point operations (FLOPs), and computational time in GPU (NVIDIA RTX 3090 24GB). The input size is $560\times560$ for ours, while other methods have the size of $512\times512$. 
\begin{table}[t]
\vspace{-5pt}
    \centering
    \scriptsize
    \setlength{\tabcolsep}{1.2pt}
    \begin{tabular}{c
        |>{\centering\arraybackslash}p{0.9cm}
        |>{\centering\arraybackslash}p{0.9cm}
        |>{\centering\arraybackslash}p{0.9cm}
        |>{\centering\arraybackslash}p{0.9cm}
        |>{\centering\arraybackslash}p{1.23cm}
        |>{\centering\arraybackslash}p{1.23cm}
        }
          &\#Params.& {Mem.} & {Time} &FLOPs & \multicolumn{2}{c}{PSNR$|$PSNR-B (dB)} \\
        \cline{1-1} \cline{6-7}
         Method &(M)&(GB)&(ms)& (G) &$q=10$&$q=40$\\ 
        \hline\hline
         FBCNN \cite{fbcnn}   & 70.1          & \cellcolor[HTML]{EFEFEF}\textbf{0.61} & {71.95}       &709.97           & 32.18$|$32.15               &36.02$|$35.95\\
         Swin2SR \cite{conde2022swin2sr}  & 11.5$\leq$ & 2.79 &2203.59&3301.5&\textcolor{black}{32.46}$|$\quad  -\quad\quad\quad &\textcolor{black}{36.25}$|$\quad  -\quad\quad\quad \\
         JDEC& 38.9 & 1.76 & 224.79& {1006.72} &\cellcolor[HTML]{EFEFEF}\textbf{\textcolor{black}{32.55}}$|$\cellcolor[HTML]{EFEFEF}\textbf{\textcolor{black}{32.51}}&\cellcolor[HTML]{EFEFEF}\textbf{\textcolor{black}{36.37}}$|$\cellcolor[HTML]{EFEFEF}\textbf{\textcolor{black}{36.28}} \\
       JDEC-CNN$^\dagger$ & \cellcolor[HTML]{EFEFEF}\textbf{26.2} & 0.81          & \cellcolor[HTML]{EFEFEF}\textbf{56.59}& \cellcolor[HTML]{EFEFEF}\textbf{476.33} &{32.31}$|${32.27}&{36.19}$|${36.09}\\
    \end{tabular}
    \vspace*{-6pt}
    \caption{Computational resources \& performance comparison for a $560\times 560$ pixels in ICB \cite{ICBdataset}. $\dagger: $We replace the deep feature extractor of \cref{Figmain} with a CNN structure for comparison with the CNN-based model \cite{fbcnn}. }
    \label{tab:model_complexity}
    \vspace{-20pt}
\end{table}

\section{Conclusion}

We proposed a local implicit neural representation approach for decoding compressed color JPEG files. Our JPEG Decoder with Enhanced Continuous cosine coefficients (JDEC) contains a novel continuous cosine formulation (CCF) to extract a high-quality spectrum of images. JDEC takes a distorted spectrum as an input of the network and decodes it to a high-quality image regardless of the given quality factor. The suggested CCF extracts the dominant components of the ground truth spectrum, effectively. The results of benchmark datasets demonstrate that our network outperforms existing models as a practical JPEG decoder.

\clearpage
\newpage
{
    \small
    \bibliographystyle{ieeenat_fullname}
    \bibliography{main}
}
\clearpage
\newpage

\appendix

\addcontentsline{toc}{section}{Appendices}
\section*{Appendix}
\setcounter{figure}{0}
\setcounter{equation}{0}
\setcounter{table}{0}

\renewcommand{\theequation}{A\arabic{equation}}
\renewcommand{\thefigure}{A\arabic{figure}}
\renewcommand{\thetable}{A\arabic{table}}

\section{Data Processing Inequality in JPEG} \label{sec:supp_dpi} In terms of information theory \cite{informationtheorybook}, it's feasible to view JPEG compression as resembling a Markov chain as follows:
\vspace*{-5pt}
\begin{align}
    X \rightarrow Y \rightarrow \hat{X},
    \label{eq:markov}
    \vspace{-20pt}
\end{align}
where, $X$ is a symbol (Images), $Y$ is a encoded file (bit-streams) and $\hat{X}$ is decoded symbol (JPEG Images). Therefore by the data processing inequality \cite{informationtheorybook}, the mutual information $I(X;Y) := \sum_{x \in \mathcal{X}} \sum_{y \in \mathcal{Y}} p(x, y) \log \frac{p(x, y)}{p(x)p(y)}
$ between $X$ and $\hat X$ cannot exceed the information between $X$ and $Y$. In JPEG, the equality is satisfied when a conventional decoder does not provide any losses. 
\vspace*{-5pt}
\begin{equation}
    I(X;Y) \geq I(X;\hat X).
    \label{eq:dataprocessingineq}
\end{equation}
The equality is satisfied when conditional mutual information $I( X;Y | \hat X) = 0$. Theoretically, a JPEG decoder should not provide loss. However, in engineering practice, loss occurs in converting the DCT spectrum back to YCbCr by performing IDCT and rounding to values of 0 and 255. Also, depending on the YCbCr to RGB matrix, precision losses occur too. Consequently, the conventional JPEG provides losses, and \cref{eq:dataprocessingineq} does not satisfy equality.

\section{Implicit Neural Representation} \label{sec:supp_inr}The proposed JDEC is a function of block coordinates. To demonstrate this, the model should be able to produce results when provided with coordinates that were not observed during training. Therefore, we conducted simple super-resolution experiments using the pre-trained JDEC to represent unobserved coordinates. We maintain all settings identical, with the size of block coordinate $\delta$ increased to $rB\times rB$ where $r$ is the upsampling ratio. 


\begin{table}[ht]
\vspace{-10pt}
\centering
\setlength{\tabcolsep}{1.2pt}
\footnotesize
\begin{tabular}{
c|>{\centering\arraybackslash}p{1.53cm}
|>{\centering\arraybackslash}p{1.53cm}
|>{\centering\arraybackslash}p{1.53cm}
|>{\centering\arraybackslash}p{1.53cm}
}
\hline
\multicolumn{5}{c}{ q=10, $\downarrow\times4$}\\
\hline
Concept & SR w/ JAR & SR w/ o JAR & SR w/ JAR & Decoding\\
\hline
Method &RRDB\cite{RRDB_ESRGAN}&SwinIR\cite{liang2021swinir}&HST\cite{li2022hst}&JDEC (\textit{ours}) \\
\hline\hline
  Set5  & 22.36  & 22.45  & \textcolor{red}{22.49}  & \textcolor{blue}{22.48} \\

\end{tabular}
\vspace*{-6pt}
\caption{\textbf{Quantitative comparison} for compressed image upsampling on the Set5 \cite{set5dataset} dataset (PSNR(dB)). The JAR refers to JPEG Artifact Removal.}
\vspace{-10pt}\label{tab:table_upsample}
\end{table}

\begin{figure}[ht]
\footnotesize
\centering
\stackunder[2pt]{\includegraphics[width = 0.7in]{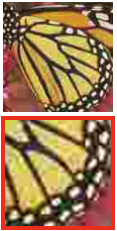}}{Input $(q=10)$}
\stackunder[2pt]{\includegraphics[width = 0.7in]{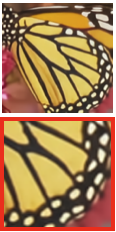}}{JDEC + Bicubic}
\stackunder[2pt]{\includegraphics[width = 1.4in]{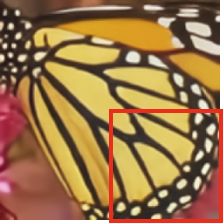}}{JDEC w/o Bicubic }
\vspace{-10pt}
\caption{\textbf{$\times2$ upsampling of compressed image ($q =10$)} in Set-5 dataset. Note that JDEC upscales the image with the change of additional coordinates $\delta$ without additional training.}
\vspace*{-5pt}
\label{fig:upsample_dis}
\vspace{-5pt}
\end{figure}
\noindent\cref{fig:upsample_dis} shows that our JDEC is clearly an implicit neural representation by extracting unseen coordinates. It also demonstrates that the resolution of estimated spectra by CCF is a function of continuous frequency. We compare our upsampling results to existing networks that aim to upsample compressed images in \cref{tab:table_upsample}. 
\section{Computational Costs and Performance}
\begin{figure}[ht]
    \footnotesize
    \centering
    \vspace{-5pt}
    \includegraphics[trim={0 0 0 0},clip,width=3.2in]{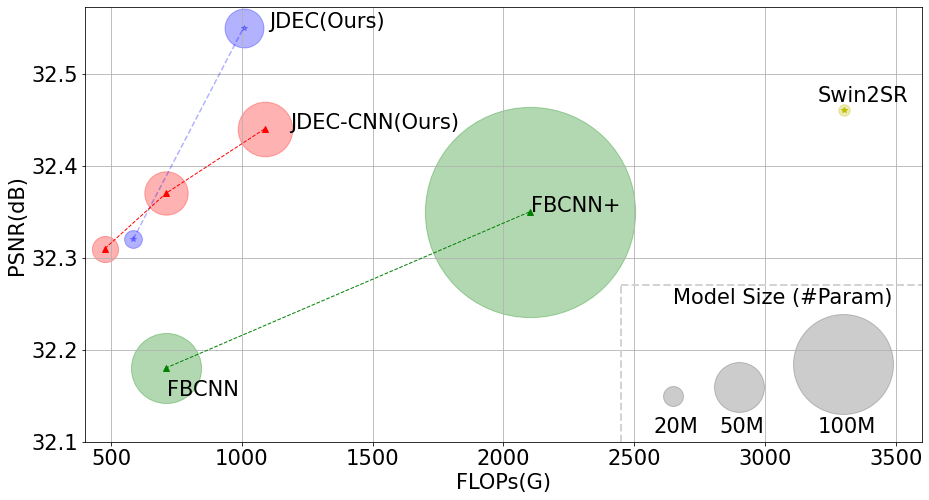}
    \vspace{-15pt}
    \caption{FLOPs and PSNR Comparison with other methods in ICB\cite{ICBdataset} ($q=10$).}
    \vspace{-15pt}
    \label{fig:complexityvsperformance}
\end{figure}

With \cref{fig:complexityvsperformance} and \cref{tab:comp_supple}, we present an additional comparison of the computational resources including an extended FBCNN model (FBCNN+) and JDEC-CNN+. Our framework overcomes the trade-off between computational complexities and performances.

\begin{table}[h]
\vspace{-10pt}
    \centering
    \scriptsize
    \setlength{\tabcolsep}{1.2pt}
    \begin{tabular}{c
        |>{\centering\arraybackslash}p{0.9cm}
        |>{\centering\arraybackslash}p{0.9cm}
        |>{\centering\arraybackslash}p{0.9cm}
        |>{\centering\arraybackslash}p{0.9cm}
        |>{\centering\arraybackslash}p{1.23cm}
        |>{\centering\arraybackslash}p{1.23cm}
        }
          &\#Params.& {Mem.} & {Time} &FLOPs & \multicolumn{2}{c}{PSNR$|$PSNR-B (dB)} \\
        \cline{1-1} \cline{6-7}
         Method &(M)&(GB)&(ms)& (G) &$q=10$&$q=40$\\ 
        \hline\hline
         FBCNN \cite{fbcnn}  & 70.1          & \cellcolor[HTML]{EFEFEF}\textbf{0.61} & {71.95}       &709.97           & 32.18$|$32.15               &36.02$|$35.95\\
              JDEC-CNN & \cellcolor[HTML]{EFEFEF}\textbf{26.2} & 0.81          & \cellcolor[HTML]{EFEFEF}\textbf{56.59}& \cellcolor[HTML]{EFEFEF}\textbf{476.33} &\cellcolor[HTML]{EFEFEF}\textbf{32.31}$|$\cellcolor[HTML]{EFEFEF}\textbf{32.27}&\cellcolor[HTML]{EFEFEF}\textbf{36.19}$|$\cellcolor[HTML]{EFEFEF}\textbf{36.09}\\
         \hline
         \hline
         FBCNN$+^\dagger$ \cite{fbcnn}& 210.3 & 1.68 & 218.58 &2101.15&\textcolor{black}{32.35}$|$\textcolor{black}{32.31} &36.20$|$36.11\\
          JDEC-CNN$+^\dagger$ & \cellcolor[HTML]{EFEFEF}\textbf{54.6} & \cellcolor[HTML]{EFEFEF}\textbf{1.62} & \cellcolor[HTML]{EFEFEF}\textbf{111.65}& \cellcolor[HTML]{EFEFEF}\textbf{1086.87} &\cellcolor[HTML]{EFEFEF}\textbf{32.43}$|$\cellcolor[HTML]{EFEFEF}\textcolor{blue}{\textbf{32.39}}&\cellcolor[HTML]{EFEFEF}\textcolor{blue}{\textbf{36.30}}$|$\cellcolor[HTML]{EFEFEF}\textcolor{blue}{\textbf{36.19}}\\
        \hline
        \hline
         Swin2SR \cite{conde2022swin2sr}& 11.5$\leq$ & 2.79 &2203.59&3301.5&\textcolor{blue}{32.46}$|$\quad  -\quad\quad\quad &\textcolor{black}{36.25}$|$\quad  -\quad\quad\quad \\
         JDEC& 38.9 & \cellcolor[HTML]{EFEFEF}\textbf{1.76} & \cellcolor[HTML]{EFEFEF}\textbf{224.79}& \cellcolor[HTML]{EFEFEF}\textbf{1006.72} &\cellcolor[HTML]{EFEFEF}\textbf{\textcolor{red}{32.55}}$|$\cellcolor[HTML]{EFEFEF}\textbf{\textcolor{red}{32.51}}&\cellcolor[HTML]{EFEFEF}\textbf{\textcolor{red}{36.37}}$|$\cellcolor[HTML]{EFEFEF}\textbf{\textcolor{red}{36.28}} \\
    \end{tabular}
    \vspace*{-6pt}
    \caption{Computational resources \& performance comparison for a $560\times 560$ pixels in ICB \cite{ICBdataset}. $^\dagger: $Our implementation}
    \label{tab:comp_supple}
    \vspace{-8pt}
\end{table}

\tikzstyle{largewindow_w} = [white, line width=0.30mm]
\tikzstyle{smallwindow_w} = [white, line width=0.10mm]
\tikzstyle{largewindow_b} = [blue, line width=0.30mm]
\tikzstyle{smallwindow_b} = [blue, line width=0.10mm]
\tikzstyle{closeup_b} = [
  opacity=1.0,          
  height=1cm,         
  width=1cm,          
  connect spies, blue  
]
\tikzstyle{closeup_w} = [
  opacity=1.0,          
  height=1cm,         
  width=1cm,          
  connect spies, white  
]
\tikzstyle{closeup_w_2} = [
  opacity=1.0,          
  height=1.7cm,         
  width=1.7cm,          
  connect spies, white  
]
\tikzstyle{closeup_w_3} = [
  opacity=1.0,          
  height=1.35cm,         
  width=1.35cm,          
  connect spies, white  
]
\tikzstyle{closeup_w_4} = [
  opacity=1.0,          
  height=0.8cm,         
  width=0.8cm,          
  connect spies, white  
]
\tikzstyle{closeup_w_5} = [
  opacity=1.0,          
  height=0.6in,         
  width=0.6in,          
  connect spies, white  
]

\begin{figure*}[t]
\footnotesize
\centering
\hspace{0pt}
\raisebox{0.4in}{\rotatebox{90}{RGB}}
\begin{tikzpicture}[x=6cm, y=6cm, spy using outlines={every spy on node/.append style={smallwindow_w}}]
\node[anchor=south] (FigA) at (0,0) {\includegraphics[trim=150 0 50 200,clip,width=1.24in]{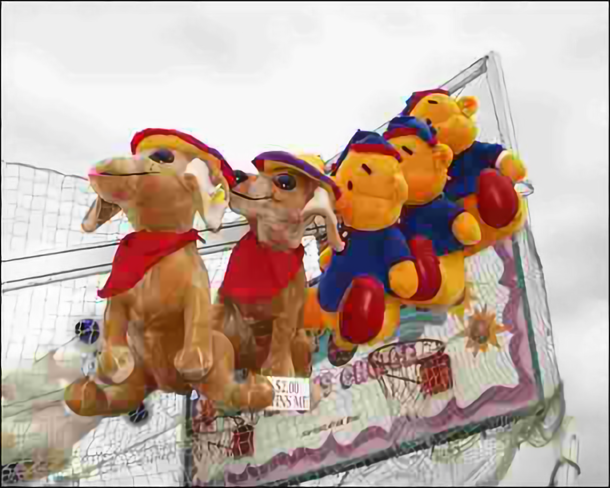}};
\spy [closeup_w,magnification=3] on ($(FigA)+(-0.080,-0.07)$)  
    in node[largewindow_w,anchor=east]      at ($(FigA.north) + ((0.26,-0.10)$); 
\spy [closeup_w,magnification=3] on ($(FigA)+(0.135,-0.047)$)  
    in node[largewindow_w,anchor=east]      at ($(FigA.north) + ((-0.1,-0.31)$); 
\end{tikzpicture}
\begin{tikzpicture}[x=6cm, y=6cm, spy using outlines={every spy on node/.append style={smallwindow_w}}]
\node[anchor=south] (FigA) at (0,0) {\includegraphics[trim=150 0 50 200,clip,width=1.24in]{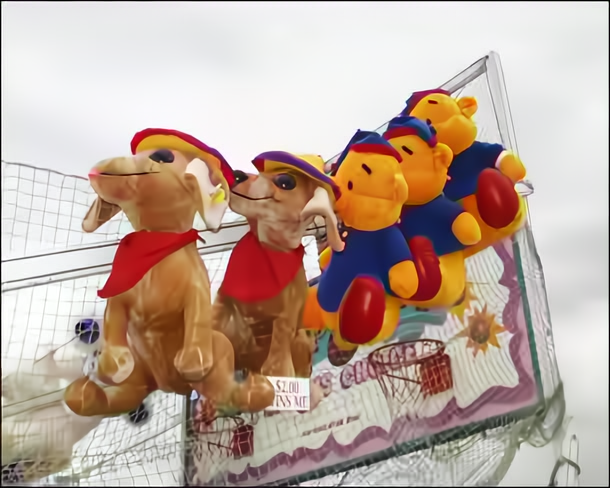}};
\spy [closeup_w,magnification=3] on ($(FigA)+(-0.080,-0.07)$)  
    in node[largewindow_w,anchor=east]      at ($(FigA.north) + ((0.26,-0.10)$); 
\spy [closeup_w,magnification=3] on ($(FigA)+(0.135,-0.047)$)  
    in node[largewindow_w,anchor=east]      at ($(FigA.north) + ((-0.1,-0.31)$); 
\end{tikzpicture}
\begin{tikzpicture}[x=6cm, y=6cm, spy using outlines={every spy on node/.append style={smallwindow_w}}]
\node[anchor=south] (FigA) at (0,0) {\includegraphics[trim=150 0 50 200,clip,width=1.24in]{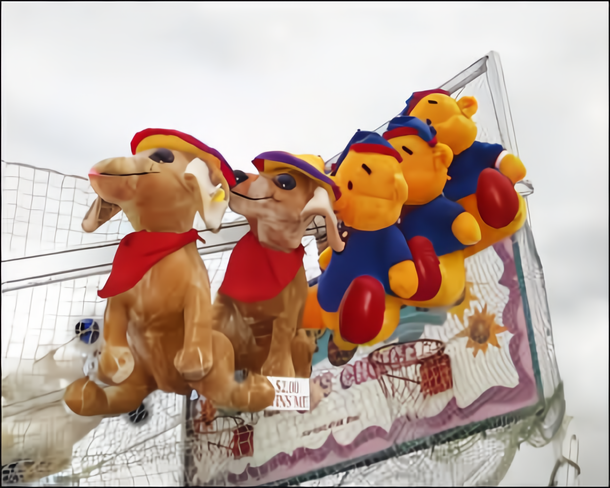}};
\spy [closeup_w,magnification=3] on ($(FigA)+(-0.080,-0.07)$)  
    in node[largewindow_w,anchor=east]      at ($(FigA.north) + ((0.26,-0.10)$); 
\spy [closeup_w,magnification=3] on ($(FigA)+(0.135,-0.047)$)  
    in node[largewindow_w,anchor=east]      at ($(FigA.north) + ((-0.1,-0.31)$); 
\end{tikzpicture}
\begin{tikzpicture}[x=6cm, y=6cm, spy using outlines={every spy on node/.append style={smallwindow_w}}]
\node[anchor=south] (FigA) at (0,0) {\includegraphics[trim=150 0 50 200,clip,width=1.24in]{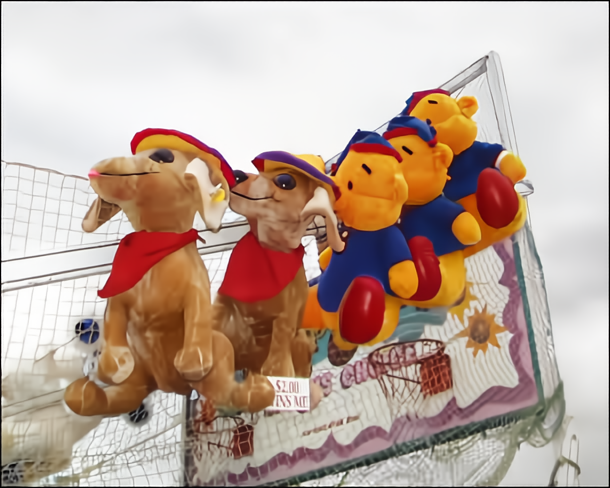}};
\spy [closeup_w,magnification=3] on ($(FigA)+(-0.080,-0.07)$)  
    in node[largewindow_w,anchor=east]      at ($(FigA.north) + ((0.26,-0.10)$); 
\spy [closeup_w,magnification=3] on ($(FigA)+(0.135,-0.047)$)  
    in node[largewindow_w,anchor=east]      at ($(FigA.north) + ((-0.1,-0.31)$); 
\end{tikzpicture}
\begin{tikzpicture}[x=6cm, y=6cm, spy using outlines={every spy on node/.append style={smallwindow_w}}]
\node[anchor=south] (FigA) at (0,0) {\includegraphics[trim=150 0 50 200,clip,width=1.24in]{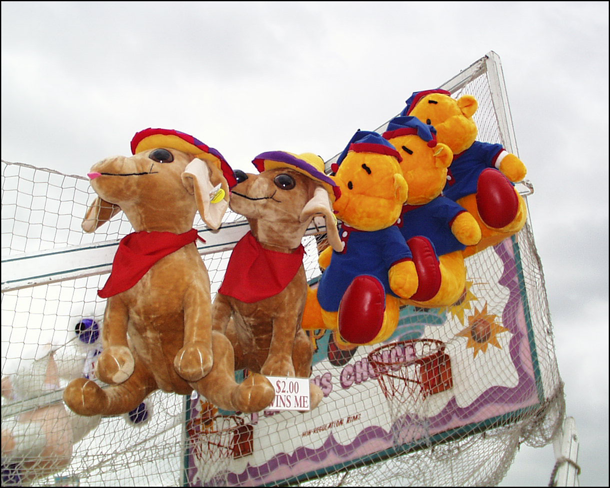}};
\spy [closeup_w,magnification=3] on ($(FigA)+(-0.080,-0.07)$)  
    in node[largewindow_w,anchor=east]      at ($(FigA.north) + ((0.26,-0.10)$); ; 
\spy [closeup_w,magnification=3] on ($(FigA)+(0.135,-0.047)$)  
    in node[largewindow_w,anchor=east]      at ($(FigA.north) + ((-0.1,-0.31)$); 
\end{tikzpicture}

\raisebox{0.4in}{\rotatebox{90}{Chroma}}
\begin{tikzpicture}[x=6cm, y=6cm, spy using outlines={every spy on node/.append style={smallwindow_w}}]
\node[anchor=south] (FigA) at (0,0) {\includegraphics[trim=150 0 50 200,clip,width=1.24in]{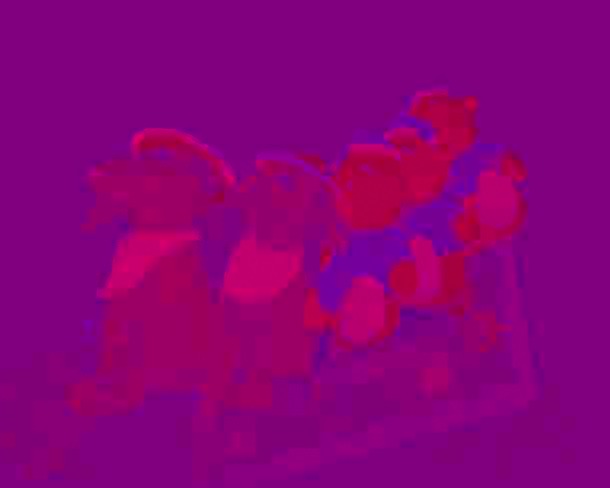}};
\spy [closeup_w,magnification=4] on ($(FigA)+(-0.080,-0.10)$)  
    in node[largewindow_w,anchor=east]      at ($(FigA.north) + ((0.26,-0.10)$); 
\spy [closeup_w,magnification=3] on ($(FigA)+(0.135,-0.047)$)  
    in node[largewindow_w,anchor=east]      at ($(FigA.north) + ((-0.1,-0.31)$); 
\end{tikzpicture}
\begin{tikzpicture}[x=6cm, y=6cm, spy using outlines={every spy on node/.append style={smallwindow_w}}]
\node[anchor=south] (FigA) at (0,0) {\includegraphics[trim=150 0 50 200,clip,width=1.24in]{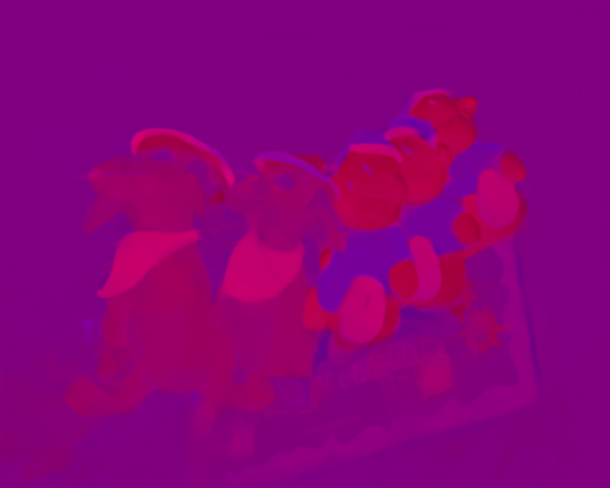}};
\spy [closeup_w,magnification=3] on ($(FigA)+(-0.080,-0.07)$)  
    in node[largewindow_w,anchor=east]      at ($(FigA.north) + ((0.26,-0.10)$); 
\spy [closeup_w,magnification=3] on ($(FigA)+(0.135,-0.047)$)  
    in node[largewindow_w,anchor=east]      at ($(FigA.north) + ((-0.1,-0.31)$); 
\end{tikzpicture}
\begin{tikzpicture}[x=6cm, y=6cm, spy using outlines={every spy on node/.append style={smallwindow_w}}]
\node[anchor=south] (FigA) at (0,0) {\includegraphics[trim=150 0 50 200,clip,width=1.24in]{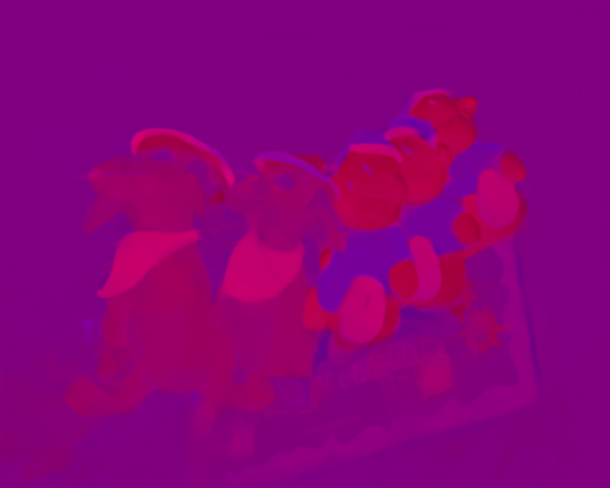}};
\spy [closeup_w,magnification=3] on ($(FigA)+(-0.080,-0.07)$)  
    in node[largewindow_w,anchor=east]      at ($(FigA.north) + ((0.26,-0.10)$); 
\spy [closeup_w,magnification=3] on ($(FigA)+(0.135,-0.047)$)  
    in node[largewindow_w,anchor=east]      at ($(FigA.north) + ((-0.1,-0.31)$); 
\end{tikzpicture}
\begin{tikzpicture}[x=6cm, y=6cm, spy using outlines={every spy on node/.append style={smallwindow_w}}]
\node[anchor=south] (FigA) at (0,0) {\includegraphics[trim=150 0 50 200,clip,width=1.24in]{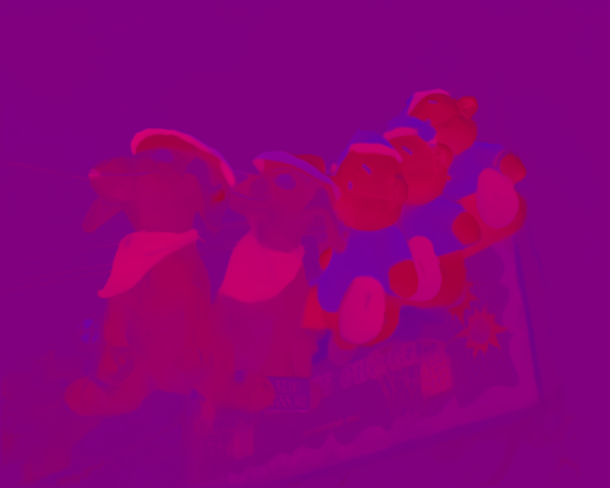}};
\spy [closeup_w,magnification=3] on ($(FigA)+(-0.080,-0.07)$)  
    in node[largewindow_w,anchor=east]      at ($(FigA.north) + ((0.26,-0.10)$); 
\spy [closeup_w,magnification=3] on ($(FigA)+(0.135,-0.047)$)  
    in node[largewindow_w,anchor=east]      at ($(FigA.north) + ((-0.1,-0.31)$); 
\end{tikzpicture}
\begin{tikzpicture}[x=6cm, y=6cm, spy using outlines={every spy on node/.append style={smallwindow_w}}]
\node[anchor=south] (FigA) at (0,0) {\includegraphics[trim=150 0 50 200,clip,width=1.24in]{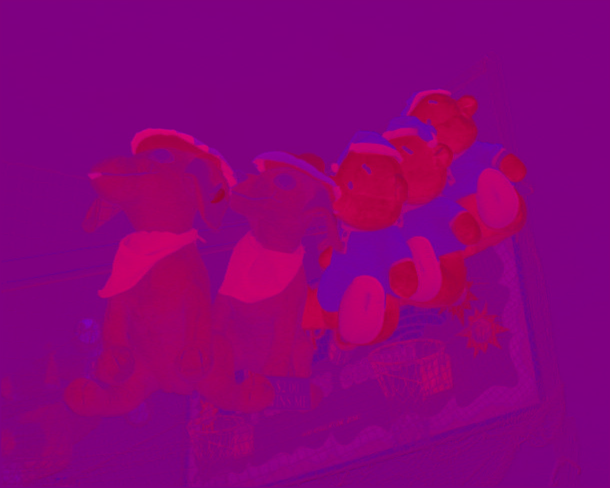}};
\spy [closeup_w,magnification=3] on ($(FigA)+(-0.080,-0.07)$)  
    in node[largewindow_w,anchor=east]      at ($(FigA.north) + ((0.26,-0.10)$); 
\spy [closeup_w,magnification=3] on ($(FigA)+(0.135,-0.047)$)  
    in node[largewindow_w,anchor=east]      at ($(FigA.north) + ((-0.1,-0.31)$); 
\end{tikzpicture}

\raisebox{0.4in}{\rotatebox{90}{RGB}}
\begin{tikzpicture}[x=6cm, y=6cm, spy using outlines={every spy on node/.append style={smallwindow_w}}]
\node[anchor=south] (FigA) at (0,0) {\includegraphics[trim=0 0 0 0,clip,width=1.24in]{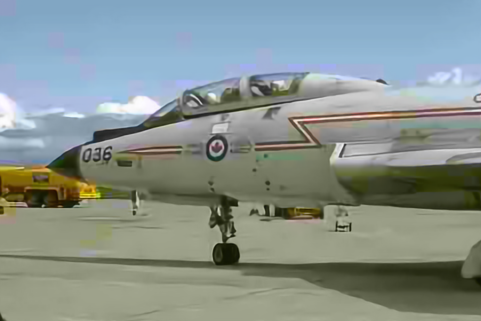}};
\spy [closeup_w,magnification=3] on ($(FigA)+(-0.210,-0.019)$)  
    in node[largewindow_w,anchor=east]      at ($(FigA.north) + ((0.26,-0.10)$); 
\spy [closeup_w,magnification=4] on ($(FigA)+(-0.02,0.01)$)  
    in node[largewindow_w,anchor=east]       at ($(FigA.north) + (0.26,-0.28)$);
\end{tikzpicture}
\begin{tikzpicture}[x=6cm, y=6cm, spy using outlines={every spy on node/.append style={smallwindow_w}}]
\node[anchor=south] (FigA) at (0,0) {\includegraphics[trim=0 0 0 0,clip,width=1.24in]{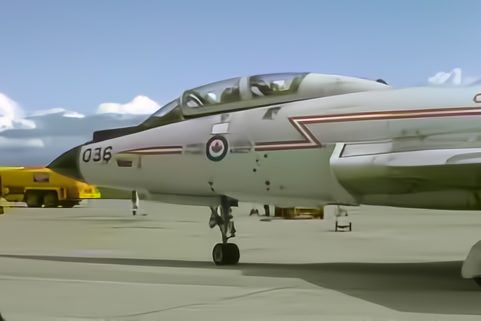}};
\spy [closeup_w,magnification=3] on ($(FigA)+(-0.210,-0.019)$)  
    in node[largewindow_w,anchor=east]      at ($(FigA.north) + ((0.26,-0.10)$); 
\spy [closeup_w,magnification=4] on ($(FigA)+(-0.02,0.01)$)  
    in node[largewindow_w,anchor=east]       at ($(FigA.north) + (0.26,-0.28)$);
\end{tikzpicture}
\begin{tikzpicture}[x=6cm, y=6cm, spy using outlines={every spy on node/.append style={smallwindow_w}}]
\node[anchor=south] (FigA) at (0,0) {\includegraphics[trim=0 0 0 0,clip,width=1.24in]{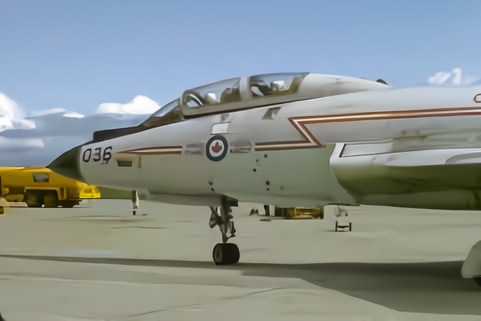}};
\spy [closeup_w,magnification=3] on ($(FigA)+(-0.210,-0.019)$)  
    in node[largewindow_w,anchor=east]      at ($(FigA.north) + ((0.26,-0.10)$); 
\spy [closeup_w,magnification=4] on ($(FigA)+(-0.02,0.01)$)  
    in node[largewindow_w,anchor=east]       at ($(FigA.north) + (0.26,-0.28)$);
\end{tikzpicture}
\begin{tikzpicture}[x=6cm, y=6cm, spy using outlines={every spy on node/.append style={smallwindow_w}}]
\node[anchor=south] (FigA) at (0,0){\includegraphics[trim=0 0 0 0,clip,width=1.24in]{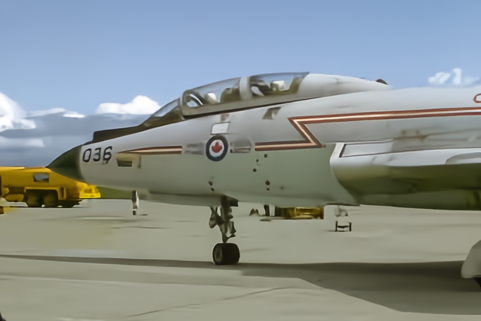}};
\spy [closeup_w,magnification=3] on ($(FigA)+(-0.210,-0.019)$)  
    in node[largewindow_w,anchor=east]      at ($(FigA.north) + ((0.26,-0.10)$); 
\spy [closeup_w,magnification=4] on ($(FigA)+(-0.02,0.01)$)  
    in node[largewindow_w,anchor=east]       at ($(FigA.north) + (0.26,-0.28)$);
\end{tikzpicture}
\begin{tikzpicture}[x=6cm, y=6cm, spy using outlines={every spy on node/.append style={smallwindow_w}}]
\node[anchor=south] (FigA) at (0,0) {\includegraphics[trim=0 0 0 0,clip,width=1.24in]{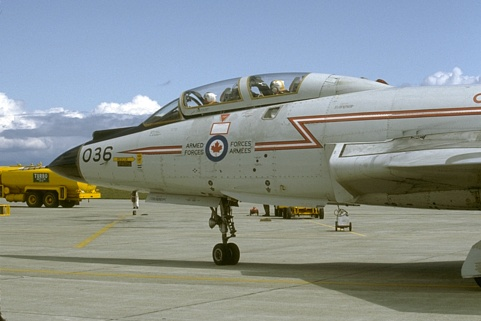}};
\spy [closeup_w,magnification=3] on ($(FigA)+(-0.210,-0.019)$)  
    in node[largewindow_w,anchor=east]      at ($(FigA.north) + ((0.26,-0.10)$); 
\spy [closeup_w,magnification=4] on ($(FigA)+(-0.02,0.01)$)  
    in node[largewindow_w,anchor=east]       at ($(FigA.north) + (0.26,-0.28)$);
\end{tikzpicture}

\hspace{-5pt}
\raisebox{0.4in}{\rotatebox{90}{Chroma}}
\begin{tikzpicture}[x=6cm, y=6cm, spy using outlines={every spy on node/.append style={smallwindow_w}}]
\node[anchor=south] (FigA) at (0,0) {\includegraphics[trim=0 0 0 0,clip,width=1.24in]{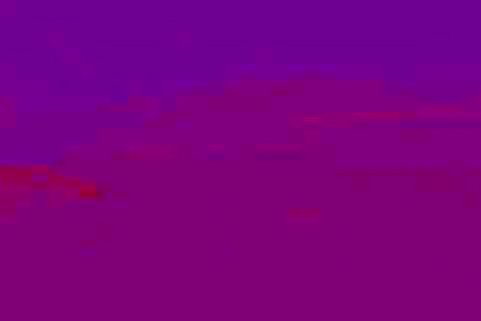}};
\spy [closeup_w,magnification=3] on ($(FigA)+(-0.210,-0.019)$)  
    in node[largewindow_w,anchor=east]      at ($(FigA.north) + ((0.26,-0.10)$); 
\spy [closeup_w,magnification=4] on ($(FigA)+(-0.02,0.01)$)  
    in node[largewindow_w,anchor=east]       at ($(FigA.north) + (0.26,-0.28)$);
\end{tikzpicture}
\begin{tikzpicture}[x=6cm, y=6cm, spy using outlines={every spy on node/.append style={smallwindow_w}}]
\node[anchor=south] (FigA) at (0,0) {\includegraphics[trim=0 0 0 0,clip,width=1.24in]{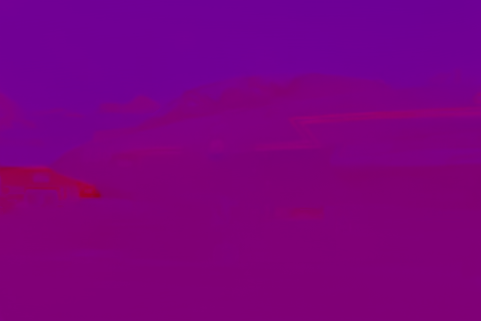}};
\spy [closeup_w,magnification=3] on ($(FigA)+(-0.210,-0.019)$)  
    in node[largewindow_w,anchor=east]      at ($(FigA.north) + ((0.26,-0.10)$); 
\spy [closeup_w,magnification=4] on ($(FigA)+(-0.02,0.01)$)  
    in node[largewindow_w,anchor=east]       at ($(FigA.north) + (0.26,-0.28)$);
\end{tikzpicture}
\begin{tikzpicture}[x=6cm, y=6cm, spy using outlines={every spy on node/.append style={smallwindow_w}}]
\node[anchor=south] (FigA) at (0,0) {\includegraphics[trim=0 0 0 0,clip,width=1.24in]{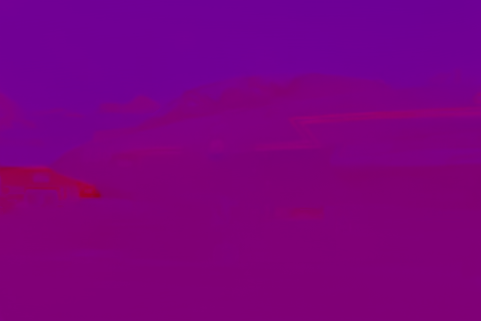}};
\spy [closeup_w,magnification=3] on ($(FigA)+(-0.210,-0.019)$)  
    in node[largewindow_w,anchor=east]      at ($(FigA.north) + ((0.26,-0.10)$); 
\spy [closeup_w,magnification=4] on ($(FigA)+(-0.02,0.01)$)  
    in node[largewindow_w,anchor=east]       at ($(FigA.north) + (0.26,-0.28)$);
\end{tikzpicture}
\begin{tikzpicture}[x=6cm, y=6cm, spy using outlines={every spy on node/.append style={smallwindow_w}}]
\node[anchor=south] (FigA) at (0,0){\includegraphics[trim=0 0 0 0,clip,width=1.24in]{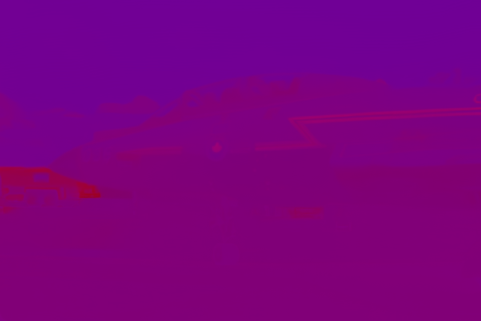}};
\spy [closeup_w,magnification=3] on ($(FigA)+(-0.210,-0.019)$)  
    in node[largewindow_w,anchor=east]      at ($(FigA.north) + ((0.26,-0.10)$); 
\spy [closeup_w,magnification=4] on ($(FigA)+(-0.02,0.01)$)  
    in node[largewindow_w,anchor=east]       at ($(FigA.north) + (0.26,-0.28)$);
\end{tikzpicture}
\begin{tikzpicture}[x=6cm, y=6cm, spy using outlines={every spy on node/.append style={smallwindow_w}}]
\node[anchor=south] (FigA) at (0,0) {\includegraphics[trim=0 0 0 0,clip,width=1.24in]{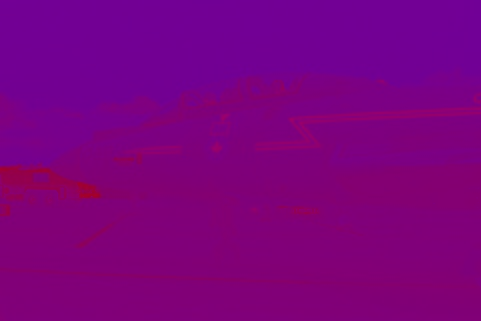}};
\spy [closeup_w,magnification=3] on ($(FigA)+(-0.210,-0.019)$)  
    in node[largewindow_w,anchor=east]      at ($(FigA.north) + ((0.26,-0.10)$); 
\spy [closeup_w,magnification=4] on ($(FigA)+(-0.02,0.01)$)  
    in node[largewindow_w,anchor=east]       at ($(FigA.north) + (0.26,-0.28)$);
\end{tikzpicture}

\hspace{-5pt}
\raisebox{0.4in}{\rotatebox{90}{RGB}}
\begin{tikzpicture}[x=6cm, y=6cm, spy using outlines={every spy on node/.append style={smallwindow_w}}]
\node[anchor=south] (FigA) at (0,0) {\includegraphics[trim=200 60 150 150,clip,width=1.24in]{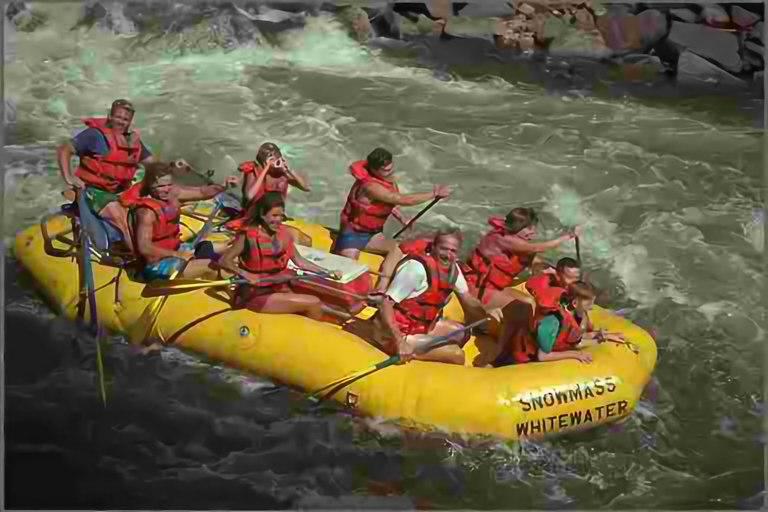}};
\spy [closeup_w,magnification=3] on ($(FigA)+(-0.175,0.075)$)  
    in node[largewindow_w,anchor=east]      at ($(FigA.north) + ((0.26,-0.10)$); 
\spy [closeup_w,magnification=3] on ($(FigA)+( 0.15, -0.15)$) 
    in node[largewindow_w,anchor=east]       at  ($(FigA.north) +((-0.1,-0.31)$);
\end{tikzpicture}
\begin{tikzpicture}[x=6cm, y=6cm, spy using outlines={every spy on node/.append style={smallwindow_w}}]
\node[anchor=south] (FigA) at (0,0) {\includegraphics[trim=200 60 150 150,clip,width=1.24in]{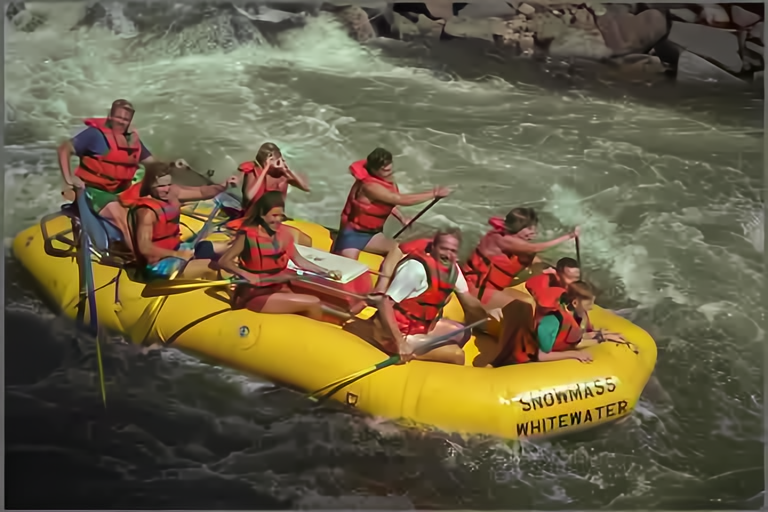}};
\spy [closeup_w,magnification=3] on ($(FigA)+(-0.175,0.075)$)  
    in node[largewindow_w,anchor=east]      at ($(FigA.north) + ((0.26,-0.10)$); 
\spy [closeup_w,magnification=3] on ($(FigA)+( 0.15, -0.15)$) 
    in node[largewindow_w,anchor=east]       at  ($(FigA.north) +((-0.1,-0.31)$);
\end{tikzpicture}
\begin{tikzpicture}[x=6cm, y=6cm, spy using outlines={every spy on node/.append style={smallwindow_w}}]
\node[anchor=south] (FigA) at (0,0) {\includegraphics[trim=200 60 150 150,clip,width=1.24in]{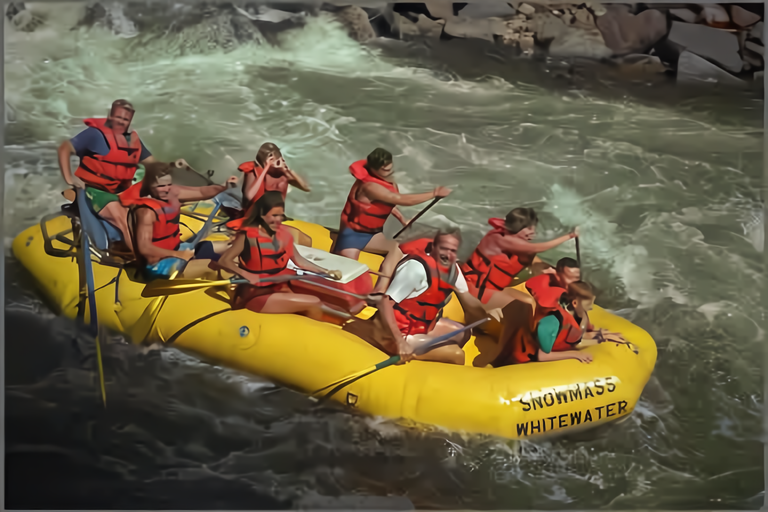}};
\spy [closeup_w,magnification=3] on ($(FigA)+(-0.175,0.075)$)  
    in node[largewindow_w,anchor=east]      at ($(FigA.north) + ((0.26,-0.10)$); 
\spy [closeup_w,magnification=3] on ($(FigA)+( 0.15, -0.15)$) 
    in node[largewindow_w,anchor=east]       at  ($(FigA.north) +((-0.1,-0.31)$);
\end{tikzpicture}
\begin{tikzpicture}[x=6cm, y=6cm, spy using outlines={every spy on node/.append style={smallwindow_w}}]
\node[anchor=south] (FigA) at (0,0) {\includegraphics[trim=200 60 150 150,clip,width=1.24in]{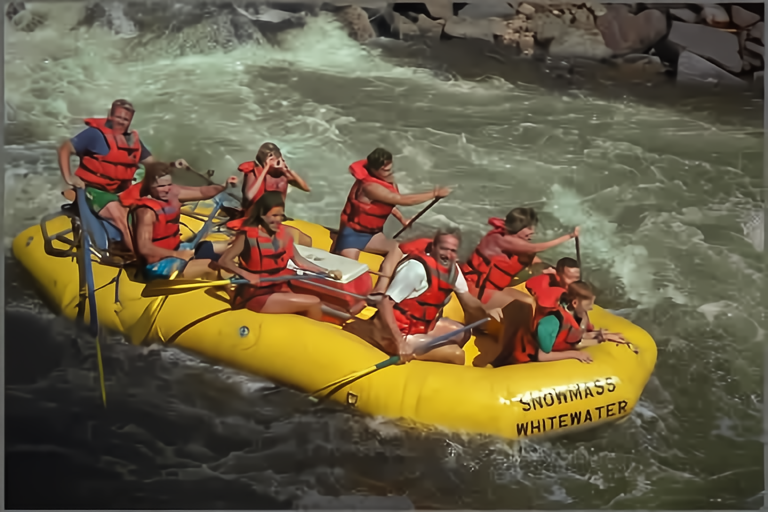}};
\spy [closeup_w,magnification=3] on ($(FigA)+(-0.175,0.075)$)  
    in node[largewindow_w,anchor=east]      at ($(FigA.north) + ((0.26,-0.10)$); 
\spy [closeup_w,magnification=3] on ($(FigA)+( 0.15, -0.15)$) 
    in node[largewindow_w,anchor=east]       at  ($(FigA.north) +((-0.1,-0.31)$);
\end{tikzpicture}
\begin{tikzpicture}[x=6cm, y=6cm, spy using outlines={every spy on node/.append style={smallwindow_w}}]
\node[anchor=south] (FigA) at (0,0) {\includegraphics[trim=200 60 150 150,clip,width=1.24in]{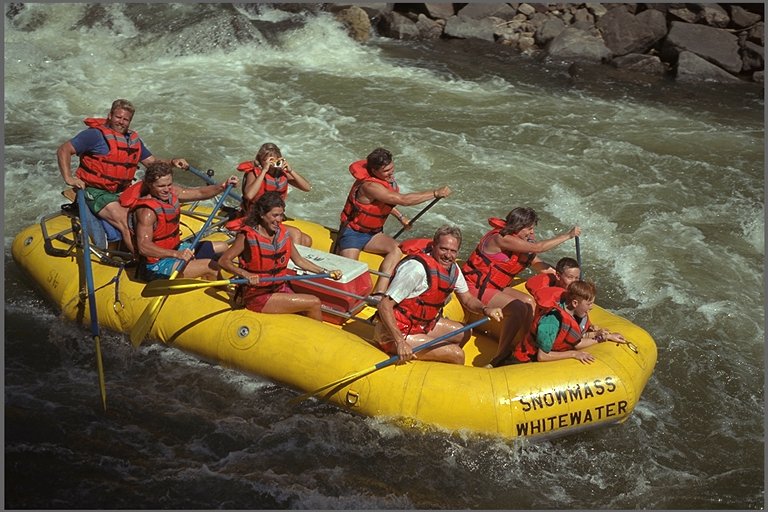}};
\spy [closeup_w,magnification=3] on ($(FigA)+(-0.175,0.075)$)  
    in node[largewindow_w,anchor=east]      at ($(FigA.north) + ((0.26,-0.10)$); 
\spy [closeup_w,magnification=3] on ($(FigA)+( 0.15, -0.15)$) 
    in node[largewindow_w,anchor=east]       at  ($(FigA.north) +((-0.1,-0.31)$);
\end{tikzpicture}

\hspace{-5pt}
\raisebox{0.4in}{\rotatebox{90}{Chroma}}
\begin{tikzpicture}[x=6cm, y=6cm, spy using outlines={every spy on node/.append style={smallwindow_w}}]
\node[anchor=south] (FigA) at (0,0) {\includegraphics[trim=200 60 150 150,clip,width=1.24in]{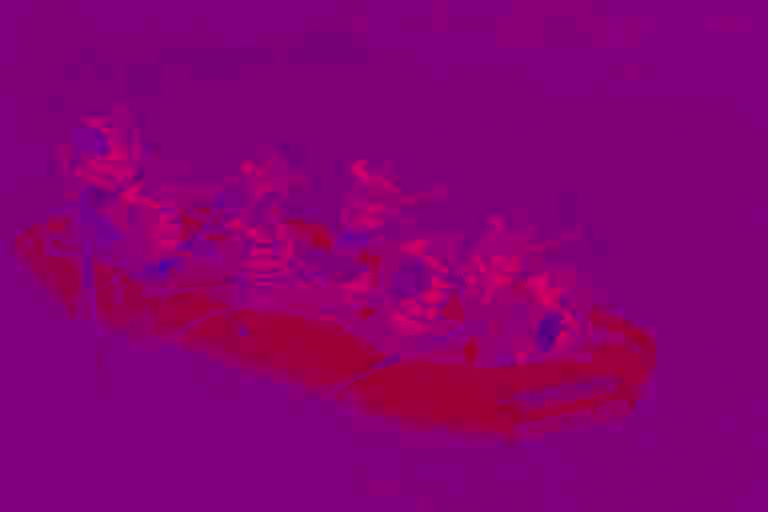}};
\spy [closeup_w,magnification=3] on ($(FigA)+(-0.175,0.075)$)  
    in node[largewindow_w,anchor=east]      at ($(FigA.north) + ((0.26,-0.10)$); 
\spy [closeup_w,magnification=3] on ($(FigA)+( 0.15, -0.15)$) 
    in node[largewindow_w,anchor=east]       at  ($(FigA.north) +((-0.1,-0.31)$);
\end{tikzpicture}
\begin{tikzpicture}[x=6cm, y=6cm, spy using outlines={every spy on node/.append style={smallwindow_w}}]
\node[anchor=south] (FigA) at (0,0) {\includegraphics[trim=200 60 150 150,clip,width=1.24in]{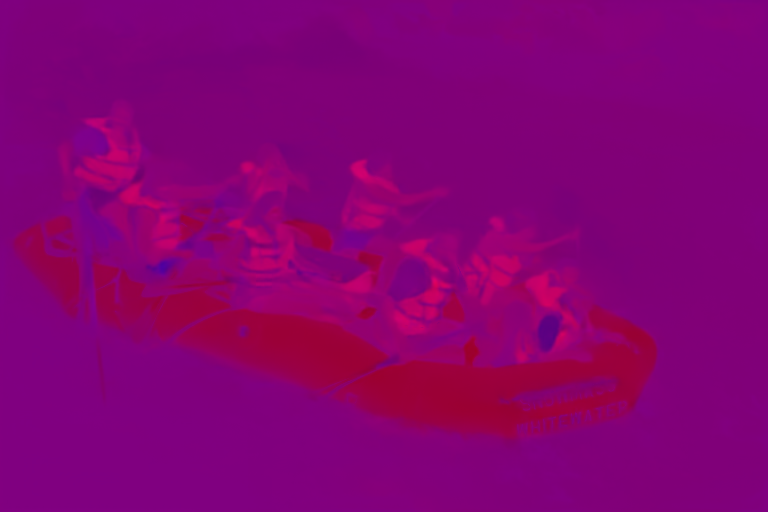}};
\spy [closeup_w,magnification=3] on ($(FigA)+(-0.175,0.075)$)  
    in node[largewindow_w,anchor=east]      at ($(FigA.north) + ((0.26,-0.10)$); 
\spy [closeup_w,magnification=3] on ($(FigA)+( 0.15, -0.15)$) 
    in node[largewindow_w,anchor=east]       at  ($(FigA.north) +((-0.1,-0.31)$);
\end{tikzpicture}
\begin{tikzpicture}[x=6cm, y=6cm, spy using outlines={every spy on node/.append style={smallwindow_w}}]
\node[anchor=south] (FigA) at (0,0) {\includegraphics[trim=200 60 150 150,clip,width=1.24in]{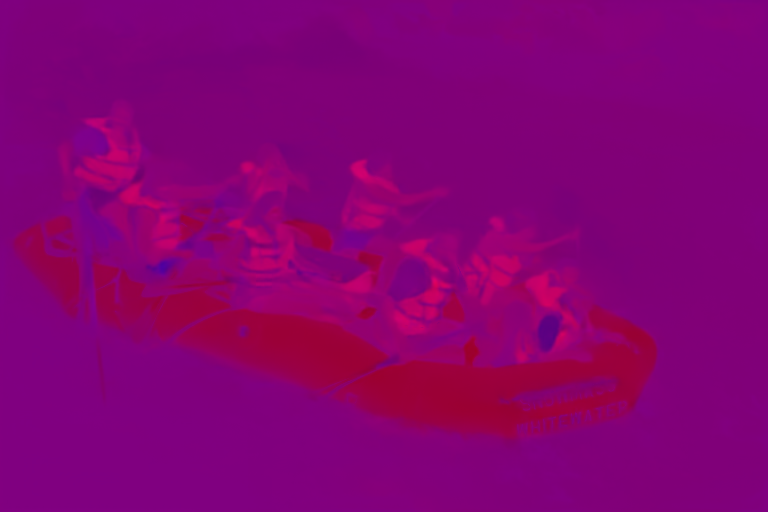}};
\spy [closeup_w,magnification=3] on ($(FigA)+(-0.175,0.075)$)  
    in node[largewindow_w,anchor=east]      at ($(FigA.north) + ((0.26,-0.10)$); 
\spy [closeup_w,magnification=3] on ($(FigA)+( 0.15, -0.15)$) 
    in node[largewindow_w,anchor=east]       at  ($(FigA.north) +((-0.1,-0.31)$);
\end{tikzpicture}
\begin{tikzpicture}[x=6cm, y=6cm, spy using outlines={every spy on node/.append style={smallwindow_w}}]
\node[anchor=south] (FigA) at (0,0) {\includegraphics[trim=200 60 150 150,clip,width=1.24in]{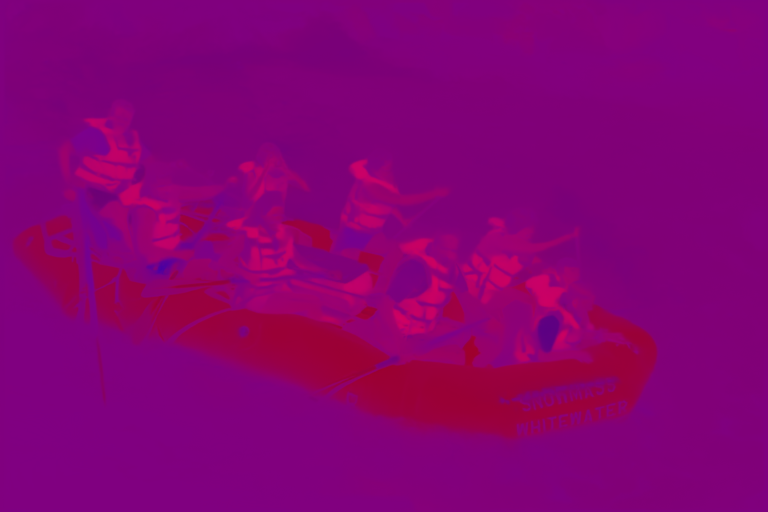}};
\spy [closeup_w,magnification=3] on ($(FigA)+(-0.175,0.075)$)  
    in node[largewindow_w,anchor=east]      at ($(FigA.north) + ((0.26,-0.10)$); 
\spy [closeup_w,magnification=3] on ($(FigA)+( 0.15, -0.15)$) 
    in node[largewindow_w,anchor=east]       at  ($(FigA.north) +((-0.1,-0.31)$);
\end{tikzpicture}
\begin{tikzpicture}[x=6cm, y=6cm, spy using outlines={every spy on node/.append style={smallwindow_w}}]
\node[anchor=south] (FigA) at (0,0) {\includegraphics[trim=200 60 150 150,clip,width=1.24in]{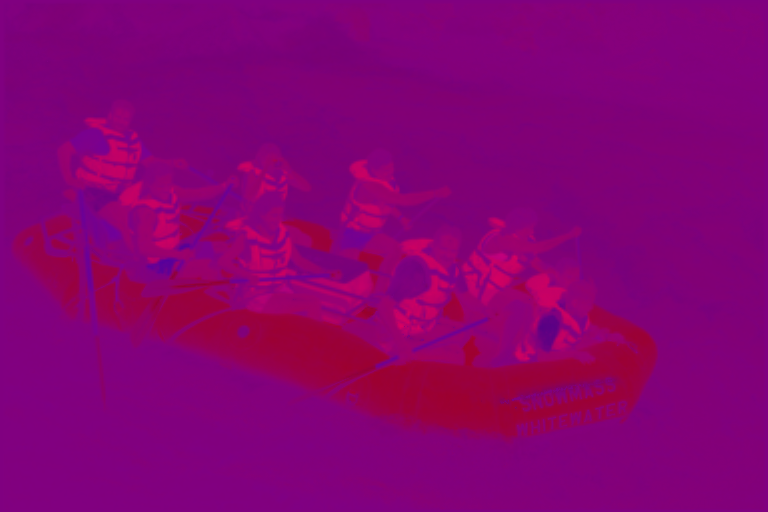}};
\spy [closeup_w,magnification=3] on ($(FigA)+(-0.175,0.075)$)  
    in node[largewindow_w,anchor=east]      at ($(FigA.north) + ((0.26,-0.10)$); 
\spy [closeup_w,magnification=3] on ($(FigA)+( 0.15, -0.15)$) 
    in node[largewindow_w,anchor=east]       at  ($(FigA.north) +((-0.1,-0.31)$);
\end{tikzpicture}

\hspace{-5pt}
\raisebox{0.4in}{\rotatebox{90}{RGB}}
\begin{tikzpicture}[x=6cm, y=6cm, spy using outlines={every spy on node/.append style={smallwindow_w}}]
\node[anchor=south] (FigA) at (0,0) {\includegraphics[trim=0 0 0 240,clip,width=1.24in]{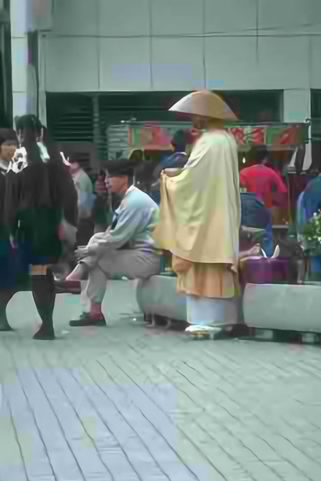}};
\spy [closeup_w,magnification=3] on ($(FigA)+( 0.10, -0.05)$) 
    in node[largewindow_w,anchor=east]       at  ($(FigA.north) +((-0.1,-0.34)$);
\end{tikzpicture}
\begin{tikzpicture}[x=6cm, y=6cm, spy using outlines={every spy on node/.append style={smallwindow_w}}]
\node[anchor=south] (FigA) at (0,0) {\includegraphics[trim=0 0 0 240,clip,width=1.24in]{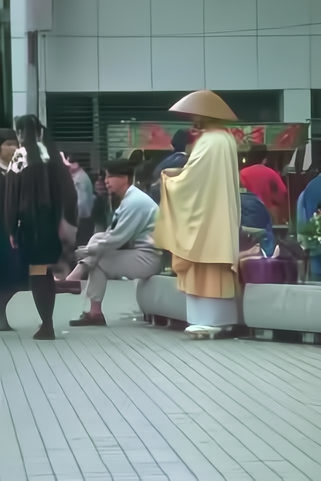}};
\spy [closeup_w,magnification=3] on ($(FigA)+( 0.10, -0.05)$) 
    in node[largewindow_w,anchor=east]       at  ($(FigA.north) +((-0.1,-0.34)$);
\end{tikzpicture}
\begin{tikzpicture}[x=6cm, y=6cm, spy using outlines={every spy on node/.append style={smallwindow_w}}]
\node[anchor=south] (FigA) at (0,0) {\includegraphics[trim=0 0 0 240,clip,width=1.24in]{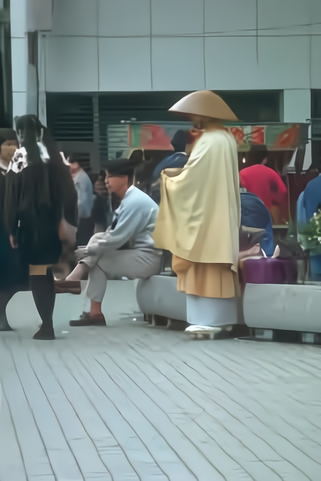}};
\spy [closeup_w,magnification=3] on ($(FigA)+( 0.10, -0.05)$) 
    in node[largewindow_w,anchor=east]       at  ($(FigA.north) +((-0.1,-0.34)$);
\end{tikzpicture}
\begin{tikzpicture}[x=6cm, y=6cm, spy using outlines={every spy on node/.append style={smallwindow_w}}]
\node[anchor=south] (FigA) at (0,0) {\includegraphics[trim=0 0 0 240,clip,width=1.24in]{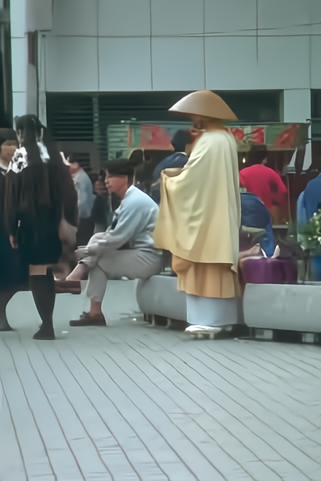}};
\spy [closeup_w,magnification=3] on ($(FigA)+( 0.10, -0.05)$) 
    in node[largewindow_w,anchor=east]       at  ($(FigA.north) +((-0.1,-0.34)$);
\end{tikzpicture}
\begin{tikzpicture}[x=6cm, y=6cm, spy using outlines={every spy on node/.append style={smallwindow_w}}]
\node[anchor=south] (FigA) at (0,0) {\includegraphics[trim=0 0 0 240,clip,width=1.24in]{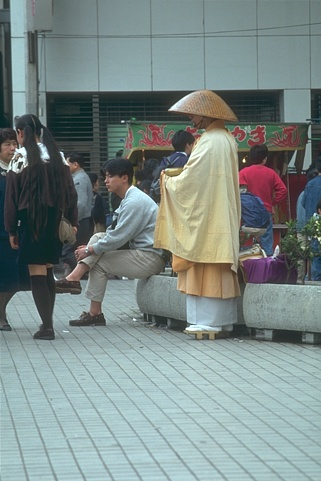}};
\spy [closeup_w,magnification=3] on ($(FigA)+( 0.10, -0.05)$) 
    in node[largewindow_w,anchor=east]       at  ($(FigA.north) +((-0.1,-0.34)$);
\end{tikzpicture}

\hspace{-5pt}
\raisebox{0.4in}{\rotatebox{90}{Chroma}}
\stackunder[2pt]{
\begin{tikzpicture}[x=6cm, y=6cm, spy using outlines={every spy on node/.append style={smallwindow_w}}]
\node[anchor=south] (FigA) at (0,0) {\includegraphics[trim=0 0 0 240,clip,width=1.24in]{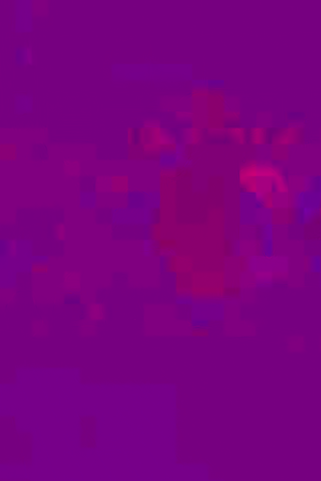}};
\spy [closeup_w,magnification=2] on ($(FigA)+(-0.180,-0.165)$)  
    in node[largewindow_w,anchor=east]      at ($(FigA.north) + ((0.26,-0.10)$); 
\end{tikzpicture}}{DnCNN \cite{dncnn}}
\stackunder[2pt]{
\begin{tikzpicture}[x=6cm, y=6cm, spy using outlines={every spy on node/.append style={smallwindow_w}}]
\node[anchor=south] (FigA) at (0,0) {\includegraphics[trim=0 0 0 240,clip,width=1.24in]{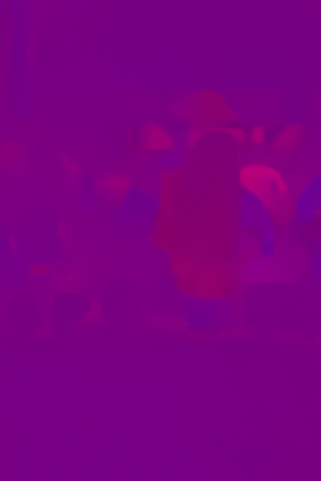}};
\spy [closeup_w,magnification=2] on ($(FigA)+(-0.180,-0.165)$)  
    in node[largewindow_w,anchor=east]      at ($(FigA.north) + ((0.26,-0.10)$); 
\end{tikzpicture}}{QGAC \cite{qgac}}
\stackunder[2pt]{
\begin{tikzpicture}[x=6cm, y=6cm, spy using outlines={every spy on node/.append style={smallwindow_w}}]
\node[anchor=south] (FigA) at (0,0) {\includegraphics[trim=0 0 0 240,clip,width=1.24in]{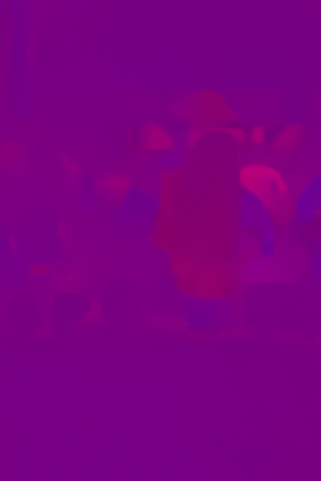}};
\spy [closeup_w,magnification=2] on ($(FigA)+(-0.180,-0.165)$)  
    in node[largewindow_w,anchor=east]      at ($(FigA.north) + ((0.26,-0.10)$); 
\end{tikzpicture}}{FBCNN \cite{fbcnn}}
\stackunder[2pt]{
\begin{tikzpicture}[x=6cm, y=6cm, spy using outlines={every spy on node/.append style={smallwindow_w}}]
\node[anchor=south] (FigA) at (0,0) {\includegraphics[trim=0 0 0 240,clip,width=1.24in]{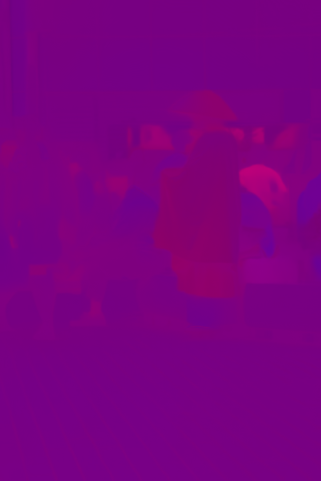}};
\spy [closeup_w,magnification=2] on ($(FigA)+(-0.180,-0.165)$)  
    in node[largewindow_w,anchor=east]      at ($(FigA.north) + ((0.26,-0.10)$); 
\end{tikzpicture}}{\textbf{JDEC \textit{(Ours)}}}
\stackunder[2pt]{
\begin{tikzpicture}[x=6cm, y=6cm, spy using outlines={every spy on node/.append style={smallwindow_w}}]
\node[anchor=south] (FigA) at (0,0) {\includegraphics[trim=0 0 0 240,clip,width=1.24in]{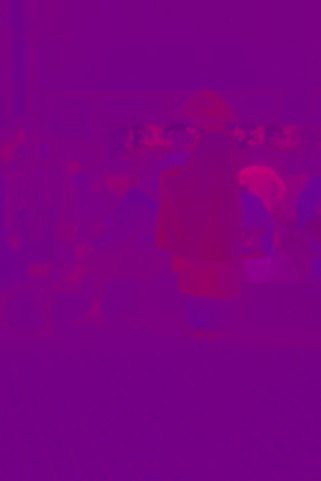}};
\spy [closeup_w,magnification=2] on ($(FigA)+(-0.180,-0.165)$)  
    in node[largewindow_w,anchor=east]      at ($(FigA.north) + ((0.26,-0.10)$); 
\end{tikzpicture}}{GT}

\vspace*{-6pt}
\caption{Additional qualitative comparison in color JPEG artifact removal ($q =10$). }
\vspace*{-12pt}
\label{fig:additional_qual}
\end{figure*}

\section{Additional Qualitative Results} 
We present additional qualitative results for comparison. Additionally, we divide RGB and chroma images to demonstrate the robustness of our JDEC to color distortion.


\clearpage
    

\end{document}